\documentclass[fleqn,usenatbib]{mnras}

\usepackage{newtxtext,newtxmath}

\usepackage[T1]{fontenc}

\DeclareRobustCommand{\VAN}[3]{#2}
\let\VANthebibliography\thebibliography
\def\thebibliography{\DeclareRobustCommand{\VAN}[3]{##3}\VANthebibliography}

\usepackage{graphicx}	
\usepackage{xcolor}
\usepackage{lastpage}
\usepackage{orcidlink}

\hypersetup{hypertexnames=false}
\usepackage{amsmath}	

\title[\(\alpha_{\rm OX}\) and disk--corona diagnostics]{Changing-look Active Galactic Nuclei from the Dark Energy Spectroscopic Instrument. VII. Testing Disk--Corona Diagnostics with eROSITA}

\author[W.-J. Guo et al.]{
Wei-Jian Guo\kern-0.3pt\orcidlink{0000-0001-9457-0589},$^{1}$\thanks{E-mail: hlhw18@durham.ac.uk; weijian.guo@durham.ac.uk}
Shuo Zhai\kern-0.3pt\orcidlink{0009-0005-4152-2088},$^{2}$
Zhu-Heng Yao\kern-0.3pt\orcidlink{0009-0000-1228-2373},$^{2}$
Huaqing Cheng\kern-0.3pt\orcidlink{0000-0003-4200-9954},$^{2}$
Jingwei Hu\kern-0.3pt\orcidlink{0000-0002-0779-1947},$^{2}$
and Jun-Jie Jin\kern-0.3pt\orcidlink{0000-0002-8402-3722},$^{2}$
\\
$^{1}$Centre for Extragalactic Astronomy, Department of Physics, Durham University, South Road, Durham, DH1 3LE, UK\\
$^{2}$National Astronomical Observatories, Chinese Academy of Sciences, 20A Datun Road, Chaoyang District, Beijing 100101, China\\
}

\date{Accepted 2026 October 5. Received 2026 September 14; in original form 2026 July 31}

\pubyear{2026}

\begin{document}
\label{firstpage}
\pagerange{\pageref{firstpage}--\pageref{LastPage}}
\maketitle


\begin{abstract}
Changing-look active galactic nuclei (CL-AGN) have been proposed to follow an X-ray-binary-like, V-shaped relation between the optical/ultraviolet-to-X-ray spectral slope, \(\alpha_{\rm OX}\), and the Eddington ratio, \(\lambda_{\rm Edd}\equiv L_{\rm bol}/L_{\rm Edd}\). We test this interpretation using CL-AGN selected from the Dark Energy Spectroscopic Instrument (DESI) and control samples matched to the eROSITA all-sky survey (eRASS). We find that the apparent positive correlation between \(\alpha_{\rm OX}\) and the optical-continuum-based Eddington ratio, \(\lambda_{\rm Edd,opt}\), is not unique to CL-AGN, but also appears in matched non-changing-look samples.  In broad-H\(\alpha\) AGN, our main statistical sample, this structure is largely driven by the known dependence of \(\alpha_{\rm OX}\) on ultraviolet luminosity and by the covariance between the rest-frame 2500~\AA\ luminosity, \(L_{2500}\), the black-hole mass, \(M_{\rm BH}\), and \(\lambda_{\rm Edd,opt}\).  When an X-ray-based Eddington ratio is used instead, the relation reverses sign because the X-ray luminosity, \(L_X\), enters the two axes with opposite signs.  An apparent \(\alpha_{\rm OX}\)--Eddington-ratio V-shape can therefore be induced by the luminosity estimator and is not a standalone accretion-state diagnostic. As a complementary test, we examine the eRASS effective photon index, \(\Gamma_{\rm eff}\), inferred from catalogue band fluxes. In broad-line AGN, it shows a weak positive association with the X-ray-independent optical Eddington ratio. However, the weaker hard-band trend depends on the null model and does not isolate coronal softening. These results motivate further X-ray spectroscopy, combined with independently estimated Eddington ratios, to test for intrinsic changes in the disk--corona system.
\end{abstract}
\begin{keywords}
galaxies: active -- galaxies: nuclei -- accretion, accretion discs -- X-rays: galaxies -- quasars: general
\end{keywords}



\section{Introduction}

Changing-look active galactic nuclei (CL-AGN) undergo large changes in their optical continuum and broad emission-line spectra on humanly accessible timescales \citep[e.g.,][]{LaMassa2015, MacLeod2019, Guo2026CLAGN}.  These events provide a rare time-domain view of the accretion flow, broad-line region, and X-ray-emitting corona in supermassive black holes \citep{NodaDone2018, RicciTrakhtenbrot2023}.  X-ray observations are particularly valuable because they probe both the innermost accretion flow and the line-of-sight absorbing material, and can therefore help distinguish intrinsic changes in the central engine from changes caused primarily by obscuration \citep{RicciTrakhtenbrot2023}.

Detailed X-ray monitoring has revealed substantial diversity among CL-AGN.  Individual objects have shown changes in the continuum slope, soft excess, reflection, absorption, and coronal emission; in extreme cases, the X-ray corona has nearly disappeared and subsequently re-formed \citep{NodaDone2018, Parker2019, Trakhtenbrot2019, Ricci2020, Ricci2021, Masterson2022, Veronese2024}.  Sample-based studies likewise indicate that optical transitions are often accompanied by intrinsic X-ray changes without the strong absorption expected from a purely obscuration-driven scenario \citep{Yang2023, Jana2025}.  More recent work has begun to follow the evolution of the photon index, soft excess, and broad-band continuum across changing states \citep{Lyu2025, Jana2026}.  These observations demonstrate that the disk and corona can evolve substantially during a changing-look event.  The central question addressed here is therefore not whether such physical changes can occur, but whether commonly used broad-band observables can identify them uniquely in a population.

To connect the heterogeneous behaviour of individual CL-AGN to a common physical framework, several studies have invoked an analogy with the accretion-state transitions of stellar-mass black-hole X-ray binaries.  In X-ray binaries, changes in accretion rate are accompanied by systematic changes in the relative disk and coronal emission and in the X-ray spectral shape \citep{RemillardMcClintock2006, Done2007}.  \citet{Sobolewska2011} scaled the spectral evolution of the Galactic black-hole binary GRO~J1655--40 to supermassive black-hole masses and predicted a change in AGN X-ray loudness near a few per cent of the Eddington luminosity.  This prediction has motivated searches for V-shaped tracks between the optical/ultraviolet-to-X-ray slope, \(\alpha_{\rm OX}\), and Eddington ratio in faded and returning changing-look quasars \citep{Ruan2019, Duffy2025, Gilbert2025}.

Interpreting these tracks requires separating physical evolution from the construction of the two axes.  The quantity \(\alpha_{\rm OX}\) measures the relative ultraviolet and X-ray normalization, not the X-ray spectral shape, and depends strongly on ultraviolet luminosity \citep{Steffen2006,Just2007,Lusso2010}.  Eddington ratios, meanwhile, can be inferred from optical/ultraviolet or X-ray luminosities through bolometric corrections \citep{Lusso2010,Lusso2012}.  Recent spectral energy distribution (SED) studies find luminosity- and accretion-dependent X-ray corrections and substantial object-to-object scatter in corrections across bands \citep{Duras2020,Gupta2024}.  Thus, luminosities entering \(\alpha_{\rm OX}\) can also enter the Eddington-ratio estimate, while the adopted correction introduces further assumptions.  Testing whether a reported morphology is specific to changing-look evolution therefore requires comparison samples selected and measured consistently.

The X-ray photon index provides a complementary diagnostic of spectral shape and its relation to accretion rate \citep{Shemmer2006,Shemmer2008,GuCao2009,Brightman2013,Yang2015}.  At low accretion rates, however, the empirical relation is not uniform across studies.  \citet{Diaz2023} showed that including reflection and high-energy coverage affects the inferred photon index and reduces scatter in its accretion-rate relation.  By contrast, \citet{Jana2023} found no photon-index--Eddington-ratio correlation in their sample of 30 AGN with \(\lambda_{\rm Edd}<10^{-3}\).  Low-luminosity AGN also require appropriate SED models rather than an untested extrapolation of luminous-quasar ultraviolet-to-X-ray and bolometric relations \citep{Lopez2024}.  These studies motivate distinguishing a spectral-shape measurement from a luminosity-based proxy.  Wide-area survey data generally lack precise source-specific spectra, but can constrain broad-band X-ray hardness statistically.  We therefore test a catalog-band hardness proxy against an accretion-rate estimate that does not use X-ray luminosity, while retaining this distinction.

In this work, we combine changing-look AGN selected from the Dark Energy Spectroscopic Instrument (DESI) \citep{DESIInstrument2022,DESIValidation2024,DESI2025} with measurements from the eROSITA all-sky survey (eRASS) \citep{Predehl2021,Sunyaev2021,Merloni2024}.  We first use the full DESI CL-AGN sample to examine their broad-band optical-to-X-ray behaviour.  We then construct a homogeneous DESI--eRASS comparison sample at \(0.35<z<0.45\), where the relevant optical emission lines are measured consistently and rest-frame 2~keV falls within the eRASS P3 band.  The comparison sample includes broad-H\(\alpha\) AGN and no-broad-H\(\alpha\) BPT-selected diagnostic samples.  With broad-H\(\alpha\) AGN as the main statistical sample, we test whether the apparent \(\alpha_{\rm OX}\)--Eddington-ratio structure is specific to CL-AGN, how it depends on the adopted luminosity estimator, and whether an eRASS catalog-band effective photon index, \(\Gamma_{\rm eff}\), provides a useful complementary diagnostic when paired with an X-ray-independent optical Eddington ratio.

The paper is organized as follows.  Sections~\ref{sec:data} and~\ref{sec:sample_measurements} describe the data, sample construction, and measurements.  Section~\ref{sec:alphaox} examines the \(\alpha_{\rm OX}\)--Eddington-ratio diagnostic in the full CL-AGN sample and in the homogeneous comparison samples.  Section~\ref{sec:gamma} tests an eRASS catalog-band X-ray-hardness diagnostic.  Section~\ref{sec:discussion} discusses the implications for CL-AGN accretion-state studies and future low-Eddington-ratio tests.  Section~\ref{sec:conclusions} summarizes our conclusions.

\section{Data}
\label{sec:data}

This section describes the optical and X-ray data used to build the uniformly matched DESI--eRASS samples.  The central design choice is to keep the optical selection, eRASS matching radius, and X-ray-band definitions fixed before comparing CL-AGN with non-changing-look controls.

\subsection{DESI spectroscopy}
\label{sec:desi}

The optical measurements are based on DESI spectra \citep{DESI2025}.  DESI covers the rest-frame optical emission-line region required to identify broad H\(\alpha\), measure narrow-line ratios for BPT classification, and estimate the optical continuum around rest-frame 5100~\AA\ in the redshift interval used for the main analysis.  We also use the DESI DR1 EmFit value-added catalogue as an external reference for initial narrow-line classification and broad-line checks.  EmFit is a low-redshift DESI emission-line fitting framework developed for measurements of Balmer, [OIII], [NII], and [SII] emission-line components, including tests for additional narrow, broad, and outflow components \citep{Pucha2025,Pucha2026}.  The final quantities used in this paper, however, are measured with our own DESI spectral fits for the eRASS-matched sample.

We refit the DESI spectra with a pPXF-based host/continuum decomposition and an AGN emission-line model \citep{Cappellari2004,Cappellari2017}.  Spectra with large internal wavelength gaps are flagged before fitting, and sources with failed or boundary fits are separated from the science classes.  For broad-H\(\alpha\) sources, we use our refitted broad-H\(\alpha\) luminosity and profile FWHM.  For no-broad-H\(\alpha\) sources, we use the stellar velocity dispersions measured from our pPXF continuum fits rather than catalogue gas-line widths or catalogue velocity-dispersion products.

\subsection{eROSITA/eRASS X-ray data}
\label{sec:erass}

The X-ray measurements come from the eROSITA all-sky survey \citep{Predehl2021} and the eRASS catalog \citep{Merloni2024}.  We use the catalog band fluxes and uncertainties as provided, and do not apply an additional source-by-source intrinsic or Galactic absorption correction.  The eRASS bands used here are P1 (\(0.2\)--\(0.5\) keV), P2 (\(0.5\)--\(1.0\) keV), P3 (\(1.0\)--\(2.0\) keV), P4 (\(2.0\)--\(5.0\) keV), and P5 (\(5.0\)--\(8.0\) keV).  For the \(\alpha_{\rm OX}\) analysis, the key band is P3.  At \(0.35<z<0.45\), rest-frame 2 keV is observed at \(1.38\)--\(1.48\) keV, so P3 provides the most direct catalog-band estimate of the monochromatic 2 keV luminosity used in \(\alpha_{\rm OX}\).

For the X-ray spectral-slope analysis, we fit an effective power-law slope to the non-overlapping P1--P5 band fluxes.  We do not include P6 (\(4.0\)--\(10.0\) keV) as an independent point because it overlaps P4 and P5.  The resulting \(\Gamma_{\rm eff}\) is therefore an eRASS catalog-band hardness proxy, not a photon index from count-level spectral fitting with source-specific response matrices and backgrounds.  This distinction is important for the interpretation in Section~\ref{sec:gamma}.

\section{Sample Construction and Measurements}
\label{sec:sample_measurements}

\subsection{Redshift window and 10 arcsec DESI--eRASS matching}
\label{sec:redshift}

The main control-sample analysis is restricted to \(0.35<z<0.45\).
This redshift window is chosen so that broad H\(\alpha\), the narrow BPT lines, and rest-frame 5100~\AA\ are measured consistently in the DESI spectra, while rest-frame 2 keV falls inside the eRASS P3 band.  We first select DESI spectra in this redshift range and then cross-match them to eRASS within a uniform radius of 10 arcsec, without preselecting on BPT class or broad-H\(\alpha\) status.

The raw 10 arcsec match contains 4300 DESI--eRASS rows, corresponding to 3652 unique DESI TARGETIDs and 3511 unique eRASS UIDs.  For the final parent sample, we deduplicate by eRASS UID, keeping the closest DESI counterpart.  This gives 3511 unique DESI--eRASS pairs.  The median separation is 4.62 arcsec, and all subsequent control-sample measurements and figures use this 10 arcsec, unique-eRASS-UID parent sample.  A 10 arcsec radius has also been used for eRASS1--Gaia matching \citep{Liu2025eRASSGaia} and nearby M-dwarf matching to stacked eRASS:4 \citep{Ilic2023eRASSMdwarfs}.  Our principal correlation trends persist when the radius is reduced to 7 arcsec (Appendix~\ref{app:matching_radius}), with only modest changes in point estimates.

The DESI and eRASS observations are not generally simultaneous.  We therefore do not interpret the matched catalog as an object-by-object evolutionary sequence.  Instead, we use it to test whether the statistical structure of the \(\alpha_{\rm OX}\)--Eddington-ratio plane is specific to CL-AGN or appears in matched non-changing-look samples selected with the same X-ray and redshift criteria.

\subsection{CL-AGN sample}
\label{sec:clagn_sample}

The parent DESI CL-AGN sample contains 561 objects selected from multi-epoch optical spectroscopy \citep{Guo2026CLAGN}.  Matching this parent sample to eRASS with the same 10 arcsec radius gives 89 catalog matches, of which 78 have finite \(\alpha_{\rm OX}\), black-hole-mass, and optical Eddington-ratio measurements.  This 10 arcsec matched CL-AGN sample provides the CL-AGN-only overview in Figure~\ref{fig:clagn10_alphaox}.

For the strict like-for-like comparison with the control samples, we also apply the \(0.35<z<0.45\) cut.  In this redshift window, the CL-AGN parent contains 54 sources; 12 have an eRASS counterpart within 10 arcsec and finite P3-based \(\alpha_{\rm OX}\) and optical Eddington-ratio measurements, and nine have a valid P1--P5 \(\Gamma_{\rm eff}\) measurement.  Because this redshift-window CL-AGN subset is small, it is used mainly as a comparison overlay and motivation, while the large control samples provide the statistical tests.

\subsection{Control-sample classes}
\label{sec:control_samples}

After fitting the 3511 unique DESI--eRASS pairs, we use two control classes.  We use the following class names throughout the paper.

Class A is the broad-H\(\alpha\) AGN sample.  These are Type~1 AGN with a detected broad H\(\alpha\) component in our refits.  This is the primary physical control sample, because both the single-epoch virial black-hole mass and the optical Eddington ratio can be estimated directly from the broad-line AGN spectrum.

Class B is the no-broad-H\(\alpha\) BPT AGN/composite sample.  These sources have no accepted broad H\(\alpha\) component, but their narrow-line ratios place them in the AGN or composite region of the BPT diagram \citep{Baldwin1981,Kewley2001,Kauffmann2003}.  They provide a narrow-line AGN control population, but their optical continuum and luminosity estimates can be more host-dominated than in Class A.

A further 196 sources are assigned to the fit-problem class and are not used in the main scientific interpretation.  The sample sizes are summarized in Table~\ref{tab:sample_summary}.

\begin{table*}
\centering
\caption{Summary of the DESI--eRASS samples used in this work.  The P3 \(\alpha_{\rm OX}\) column gives the number of sources with finite optical quantities and valid P3-based \(L_{\rm 2keV}\).  The P1--P5 \(\Gamma_{\rm eff}\) column gives the number of sources with enough valid eRASS band-flux information to fit the effective X-ray slope. Class A/B and CL-AGN counts use the independent band selection.}
\label{tab:sample_summary}
\begin{tabular}{lccc}
\hline
Sample & Parent/classified & P3 \(\alpha_{\rm OX}\) & P1--P5 \(\Gamma_{\rm eff}\) \\
\hline
DESI--eRASS parent, \(0.35<z<0.45\) & 3511 & -- & -- \\
DESI CL-AGN parent & 561 & -- & -- \\
CL-AGN/eRASS, all redshifts, 10 arcsec & 89 & 78 & -- \\
CL-AGN, \(0.35<z<0.45\), 10 arcsec & 54 & 12 & 9 \\
Class A: broad-H\(\alpha\) AGN & 2171 & 2032 & 1695 \\
Class B: no-broad-H\(\alpha\) BPT AGN/composite & 656 & 410 & 285 \\
Fit-problem class, excluded & 196 & -- & -- \\
\hline
\end{tabular}
\end{table*}

\subsection{Cross-matched analysis samples}
\label{sec:crossmatched_samples}

\subsubsection{P3 \texorpdfstring{\(\alpha_{\rm OX}\)}{alphaOX} sample}

\begin{figure*}
    \centering
    \includegraphics[width=\textwidth]{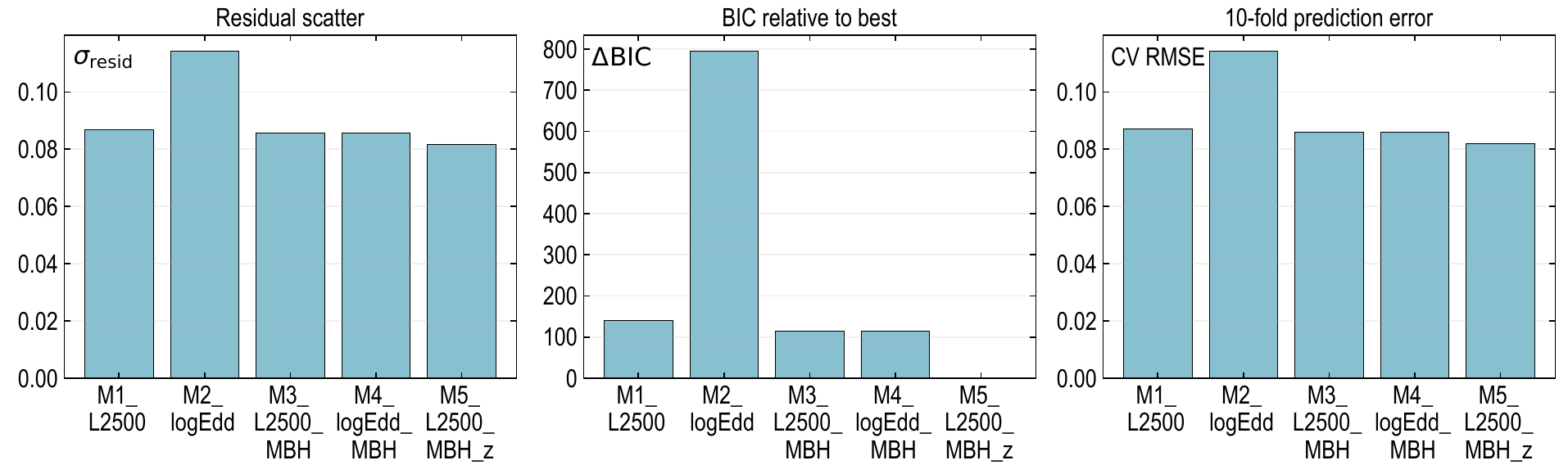}
    \caption{
    \textbf{Regression and residual tests for Class A broad-H\(\alpha\) AGN.}
    Model comparison shows that \(L_{2500}\) is a much stronger single predictor of \(\alpha_{\rm OX}\) than \(\lambda_{\rm Edd,opt}\).  Residual tests show that subtracting the \(\alpha_{\rm OX}\)--\(L_{2500}\) relation leaves only a weak mass dependence, whereas subtracting the \(\alpha_{\rm OX}\)--\(\lambda_{\rm Edd,opt}\) relation leaves a strong residual mass trend.  These results support the interpretation that the apparent \(\alpha_{\rm OX}\)--Eddington-ratio structure is largely a projection of the primary UV-luminosity dependence.
    }
    \label{fig:alphaox_regression}
\end{figure*}

The main \(\alpha_{\rm OX}\) sample requires a 10 arcsec eRASS match, finite optical quantities, and a positive P3-based estimate of rest-frame \(L_{\rm 2keV}\).  This is the sample used for Figures~\ref{fig:alphaox_lam_samples} and \ref{fig:alphaox_regression}.  For higher-significance consistency checks and for the black-hole-mass-bin panels, we additionally require \({\rm DET\_LIKE}_{\rm P3}\geq6\).
This stricter P3 likelihood cut leaves 1200 Class A sources and 211 Class B sources in the mass-bin figure.

\subsubsection{\texorpdfstring{P1--P5 \(\Gamma_{\rm eff}\)}{P1--P5 Gammaeff} sample}

The \(\Gamma_{\rm eff}\) sample is defined separately from the P3 \(\alpha_{\rm OX}\) sample.  It requires enough valid eRASS P1--P5 band fluxes and uncertainties to fit an effective power-law slope, together with finite optical Eddington-ratio and black-hole-mass estimates.  For the band-flux fit, a band must have positive flux and \({\rm DET\_LIKE}\ge3\), and at least two bands must pass this cut.  Requiring positive finite flux errors as well gives 1695 Class A objects, 285 Class B objects and nine redshift-window CL-AGN, without a parent P3 likelihood cut.  The original 1063/166 Class A/B counts refer instead to the \({\rm DET\_LIKE}_{\rm P3}\ge6\) quality subset.  The original full-band null/stacking displays are retained on their original quality selection, not reinterpreted as tests of the independent 1695-object sample.  The main \(\Gamma_{\rm eff}\) interpretation uses Class A, because its optical Eddington ratio is the cleanest X-ray-independent accretion-rate proxy.

We do not constrain the line-of-sight hydrogen column density, \(N_{\rm H}\), source by source in this catalogue-band analysis.  Absorption can suppress soft-band fluxes and harden \(\Gamma_{\rm eff}\), so the index cannot isolate intrinsic spectral changes.  We plan to present a detailed investigation of these absorption effects in a future paper based on \emph{Einstein Probe} follow-up observations.

\subsection{Derived quantities}
\label{sec:derived}

\subsubsection{\texorpdfstring{The optical/UV-to-X-ray slope \(\alpha_{\rm OX}\)}{alphaOX}}

We define the optical/UV-to-X-ray slope as \citep[e.g.,][]{Steffen2006,Just2007,Lusso2010}
\begin{equation}
\alpha_{\rm OX}
=
0.3838
\log
\left(
\frac{L_{2500}}{L_{\rm 2keV}}
\right).
\label{eq:alphaox}
\end{equation}
This sign convention is the opposite of the classical convention sometimes used in the literature.  With our definition, a larger \(\alpha_{\rm OX}\) indicates weaker X-ray emission relative to the UV.  The rest-frame 2500~\AA\ luminosity, \(L_{2500}\), is inferred from the adopted optical continuum luminosity assuming the optical/UV continuum slope of the quasar composite spectrum \citep{VandenBerk2001}. For the Class A/B control samples, we adopt \(L_\nu\propto\nu^{-0.5}\). At fixed sample membership and unchanged Eddington-ratio estimates, changing this common index shifts \(\alpha_{\rm OX}\) by a constant, preserving Spearman/Pearson coefficients and free-intercept regression slopes.

The rest-frame 2 keV luminosity, \(L_{\rm 2keV}\), is inferred from the eRASS P3-band flux.  We assume a local power-law energy spectrum,
\begin{equation}
F_E=A E^{1-\Gamma_X},
\qquad
\Gamma_X=1.8,
\label{eq:xray_powerlaw}
\end{equation}
where \(E\) is in observed-frame keV and \(F_E\) is the observed energy-flux density.  The normalization is set by the catalog P3 energy flux,
\begin{equation}
A=
\frac{F_{\rm P3}}
{\int_{1\,{\rm keV}}^{2\,{\rm keV}} E^{1-\Gamma_X}\,{\rm d}E}.
\label{eq:p3_normalization}
\end{equation}
At redshift \(z\), rest-frame 2 keV corresponds to \(E_0=2/(1+z)\) keV in the observed frame.  We convert to a monochromatic luminosity using
\begin{equation}
L_\nu(2\,{\rm keV})
=
\frac{4\pi D_L^2}{1+z}\,h\,A E_0^{1-\Gamma_X},
\label{eq:p3_lnu_2kev}
\end{equation}
where \(h\) is expressed in keV s.  The homogeneous comparison uses \(0.35<z<0.45\); the P3 conversion applies whenever \(E_0\) lies within 1--2 keV, rather than only in that window. Soft-excess emission could affect the P3 estimate; Appendix~\ref{app:m1_p4} tests a harder-band-anchored extrapolation without measuring or subtracting a soft-excess component.

The definition in Equation~\ref{eq:alphaox} makes \(\alpha_{\rm OX}\) directly sensitive to both the UV continuum and X-ray luminosity.  Therefore, any Eddington-ratio estimator based on either the optical/UV continuum or X-ray luminosity will be covariant with \(\alpha_{\rm OX}\).  This covariance is the central motivation for the tests in Section~\ref{sec:alphaox}.

\subsubsection{Optical continuum, black-hole mass, and [OIII] calibration}

For Class A broad-H\(\alpha\) AGN, we infer \(L_{5100}\) from our measured broad-H\(\alpha\) luminosity using the empirical H\(\alpha\)--continuum relation \citep{GreeneHo2005}.  Here \(L_{5100}\) is inferred from the measured broad-H\(\alpha\) luminosity; this inferred continuum luminosity and the measured broad-H\(\alpha\) profile FWHM enter the adopted mass prescription below. Our inferred \(L_{5100}\) depends on the broad-H\(\alpha\) luminosity through the tight empirical line--continuum relation of \citet{GreeneHo2005}, which retains finite scatter. We then estimate \(M_{\rm BH}\) from the profile FWHM of the fitted broad-H\(\alpha\) component and \(L_{5100}\) using a single-epoch virial form \citep[e.g.,][]{VestergaardPeterson2006},
\begin{equation}
\log \left(\frac{M_{\rm BH}}{M_\odot}\right)
=
6.91
+0.5\log\left(\frac{L_{5100}}{10^{44}{\rm erg~s^{-1}}}\right)
+2\log\left(\frac{{\rm FWHM}_{\rm H\alpha}}{10^3{\rm km~s^{-1}}}\right).
\label{eq:mbh_virial}
\end{equation}
The broad-H\(\alpha\) line uncertainties used in the plots are derived from the emission-line refits and Monte Carlo perturbations of the fitted line model.

For Class B, no broad H\(\alpha\) component is accepted.  We therefore estimate \(M_{\rm BH}\) from stellar velocity dispersions measured with our pPXF continuum fits \citep{Cappellari2004,Cappellari2017}, requiring the fitted dispersion to be non-boundary and to satisfy \(80<\sigma_*<500~{\rm km~s^{-1}}\).  We then apply the \(M_{\rm BH}\)--\(\sigma_*\) calibration
\begin{equation}
\log \left(\frac{M_{\rm BH}}{M_\odot}\right)
=
8.49
+4.38\log\left(\frac{\sigma_*}{200~{\rm km~s^{-1}}}\right)
\label{eq:mbh_sigma}
\end{equation}
from \citet{KormendyHo2013}.  We do not use [OIII] gas-line width as the fiducial black-hole-mass estimator, because it is a noisy proxy for the stellar potential in this sample.

For our Class A spectra, we favour broad-H\(\alpha\) virial masses as the more reliable choice, given potential nuclear continuum dilution of stellar absorption features \citep{GreeneHo2006}.  For Class B, which lacks accepted broad H\(\alpha\), we instead use the measured stellar velocity dispersion and the \(M_{\rm BH}\)--\(\sigma_*\) relation.

For Class B, the optical luminosity scale must be inferred from narrow-line emission.  We calibrate [OIII]-based bolometric luminosities against the Class A/broad-H\(\alpha\) scale using a DESI QSO calibration sample at \(0.35<z<0.45\).  The broad-H\(\alpha\)-based luminosity is taken as \(L_{\rm bol,H\alpha}=9L_{5100}\), using the fixed approximation \(k_{5100}=9\) \citep{Kaspi2000}, subject to object-dependent SED differences \citep{Richards2006}.  We then adopt
\begin{equation}
L_{\rm bol,[OIII]}=k_{\rm [OIII]}L_{\rm [OIII]},
\label{eq:oiii_lbol}
\end{equation}
with \(k_{\rm [OIII]}=1.79\times10^3\).  This is an empirical internal calibration that places the no-broad-H\(\alpha\) controls on the same luminosity scale as the broad-H\(\alpha\) sample; it is not intended as a universal physical bolometric correction.  It replaces the conventional uncalibrated \(k=3500\) scale \citep{Heckman2004}, which is high by \(0.29\) dex for our calibration sample.  The residual scatter after calibration is \(0.33\) dex, so the [OIII]-based luminosities are used only as a statistical scale for the no-broad-H\(\alpha\) controls.  The calibration is shown in Appendix~\ref{app:oiii_calibration}.

\subsubsection{Optical Eddington ratio}

Our fiducial Eddington ratio is based on the optical continuum:
\begin{equation}
\lambda_{\rm Edd,opt}
=
\frac{L_{\rm bol,opt}}{L_{\rm Edd}}
=
\frac{k_{5100}L_{5100}}{L_{\rm Edd}},
\label{eq:lambda_opt}
\end{equation}
where \(k_{5100}=9\) is fixed for every relevant object, not luminosity- or Eddington-ratio-dependent \citep{Kaspi2000}; SED diversity remains a systematic limitation \citep{Richards2006}, and
\begin{equation}
L_{\rm Edd}
=
1.26\times10^{38}
\left(
\frac{M_{\rm BH}}{M_\odot}
\right)
{\rm erg~s^{-1}} .
\label{eq:ledd}
\end{equation}
The masses entering Equation~\ref{eq:ledd} are estimated using Equations~\ref{eq:mbh_virial} and~\ref{eq:mbh_sigma} for Classes A and B, respectively.
The estimator in Equation~\ref{eq:lambda_opt} is independent of the eRASS X-ray flux and is therefore our preferred accretion-rate proxy for the \(\Gamma_{\rm eff}\) analysis.  However, it is not independent of \(\alpha_{\rm OX}\), because \(L_{2500}\) is inferred from the same optical continuum.  This point is examined explicitly in Section~\ref{sec:alphaox}.

\subsubsection{X-ray-based Eddington ratio}

For systematic tests, we also define an X-ray-based Eddington ratio using the rest-frame \(2\)--\(10\) keV luminosity and an X-ray bolometric correction \citep[e.g.,][]{Lusso2012}:
\begin{equation}
\lambda_{\rm Edd,X}
=
\frac{k_XL_{2-10}}{L_{\rm Edd}},
\label{eq:lambda_x}
\end{equation}
or equivalently use \(L_{2-10}/L_{\rm Edd}\) as an X-ray Eddington-scaled luminosity. Here \(k_X=20\) is fixed for all relevant objects.  Fixed alternatives 10 and 30 merely translate \(\log\lambda_{\rm Edd,X}\) for the same objects and do not change their rank correlation or fitted slope (Appendix~\ref{app:m3_kbol}).  For the \(\alpha_{\rm OX}\) estimator-dependence test, \(L_{2-10}\) is obtained from the P3-based \(L_\nu(2\,{\rm keV})\) (Equation~\ref{eq:p3_lnu_2kev}) using the same \(\Gamma_X=1.8\) power law.  For the \(\Gamma_{\rm eff}\) systematics test, we also compute a catalog hard-band version by fitting only the normalization to P4/P5 with \(\Gamma_X=1.8\), and a coupled version by integrating the best-fitting P1--P5 model over observed-frame \(2/(1+z)\)--\(10/(1+z)\) keV.  This quantity is not used as the fiducial Eddington ratio for the \(\alpha_{\rm OX}\) analysis, because \(L_X\) enters \(\alpha_{\rm OX}\) and \(\lambda_{\rm Edd,X}\) in opposite directions.  It is instead used to demonstrate how an apparent negative branch can be induced when the horizontal axis is X-ray based. We note that the optical- and X-ray-based Eddington-ratio estimates need not agree.

\subsubsection{\texorpdfstring{Effective X-ray photon index \(\Gamma_{\rm eff}\)}{Gammaeff}}

We estimate an effective X-ray photon index, \(\Gamma_{\rm eff}\), from the eRASS band fluxes by fitting a power-law spectral shape to the available eRASS bands.  The primary \(\Gamma_{\rm eff}\) measurement uses the non-overlapping P1--P5 band fluxes.  For each band \(i\), the model flux is
\begin{equation}
F_i=N_X\int_{E_{i,1}}^{E_{i,2}}E^{1-\Gamma_{\rm eff}}\,{\rm d}E .
\label{eq:band_flux_model}
\end{equation}
We fit Equation~\ref{eq:band_flux_model} in log-flux space, with \(N_X\) free for every trial \(\Gamma_{\rm eff}\).  The weights are based on the catalog fractional flux errors, \(\sigma_{\ln F}\simeq\sigma_F/F\), clipped to avoid a single band dominating the fit.  The quoted \(\Gamma_{\rm eff}\) uncertainty is the \(\Delta\chi^2=1\) interval from the one-parameter grid after profiling analytically over the normalization.  The grid is \(0\le\Gamma_{\rm eff}\le5\) in steps of 0.001 and the fractional-error weights are clipped to 0.05--5.  Catalogue fluxes inherit a reference spectral model (\(\Gamma=2\)) through their energy-conversion factors; this is distinct from the post-processing \(\Gamma_X=1.8\) and mock input \(\Gamma_{\rm true}=1.9\).  Thus \(\Gamma_{\rm eff}\) remains a model-dependent band-flux index, not a response-aware photon-index measurement.

The \(\Gamma_{\rm eff}\) analysis is interpreted statistically.  We use \(\Gamma_{\rm eff}\) to test whether the average X-ray spectral hardness depends on an X-ray-independent optical Eddington ratio.  We also perform null simulations, hard-band-selected subsample tests, and eRASS band-flux stacking to assess whether any observed trend can be produced by the band-flux fitting procedure or by the eRASS detection selection.  For the retained original P3-quality-selected full-band null test, each source keeps its observed redshift, optical luminosity, black-hole mass, optical Eddington ratio, catalog band-flux errors, and DET\_LIKE values.  We set \(\Gamma_{\rm true}=1.9\), normalize the mock spectrum to the observed P3 flux, generate P1--P5 fluxes, perturb them with Gaussian noise using the catalog flux errors, reapply the same P3 and per-band detection cuts, and run the same P1--P5 fitting pipeline.  The final null distribution uses 1000 mock realizations. DET\_LIKE values are held fixed, not regenerated from photon counts.  For the independent sample we additionally fit P3--P5 alone and perform the three conditional hard-band null tests in Appendix~\ref{app:m2_hard}, retaining the same flux-fitting pipeline.

\section{Disk--Corona Coupling from \texorpdfstring{\(\alpha_{\rm OX}\)}{alphaOX}}
\label{sec:alphaox}

In this section we test whether the \(\alpha_{\rm OX}\)--Eddington-ratio plane provides an independent diagnostic of disk--corona coupling in CL-AGN.  We first examine the apparent relation between \(\alpha_{\rm OX}\) and the optical Eddington ratio in the CL-AGN and in the matched control samples.  We then show that the observed trends and black-hole-mass stratification are largely expected from the shared dependence on the optical/UV continuum luminosity.  Next, we demonstrate that changing the Eddington-ratio estimator can reverse the sign of the apparent relation, producing an estimator-induced branch that resembles part of a V-shaped morphology.  We close the section with multivariate regressions that identify which variables actually drive the observed \(\alpha_{\rm OX}\) structure.

\subsection{The apparent \texorpdfstring{\(\alpha_{\rm OX}\)--optical Eddington-ratio relation}{alphaOX--optical Eddington-ratio relation}}
\label{sec:alphaox_optical}

\begin{figure}
    \centering
    \includegraphics[width=\columnwidth]{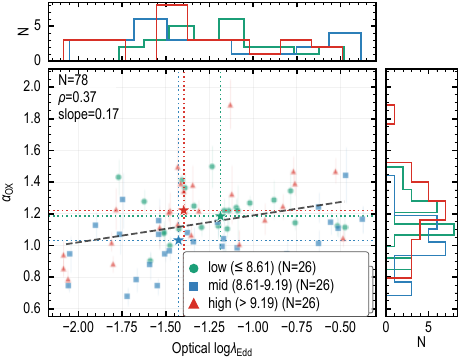}
    \caption{
    \textbf{CL-AGN-only \(\alpha_{\rm OX}\)--optical Eddington-ratio overview.}
    The eRASS-matched DESI CL-AGN sample contains 89 catalog matches within 10 arcsec; 78 have finite optical Eddington ratios, black-hole masses, and catalog-band rest-frame 2 keV luminosities.  The source colour indicates black-hole-mass tertile, and star symbols mark the bin medians.  Faint vertical error bars show the propagated uncertainty from the eRASS flux entering \(\alpha_{\rm OX}\).  This figure motivates the matched-control test: an apparent positive relation is visible in CL-AGN, but its physical interpretation requires comparison to non-changing-look samples selected with the same X-ray and redshift criteria.
    }
    \label{fig:clagn10_alphaox}
\end{figure}

We first examine the CL-AGN sample by itself.  Figure~\ref{fig:clagn10_alphaox} shows the 78 eRASS-matched CL-AGN within 10 arcsec that have finite \(\alpha_{\rm OX}\), optical Eddington ratio, and black-hole mass under the same optical-luminosity prescription used for the control analysis.  This CL-AGN-only view shows an apparent positive trend between \(\alpha_{\rm OX}\) and \(\lambda_{\rm Edd,opt}\), with \(\rho=0.37\) and a fitted slope of \(0.17\).  The point of the rest of the section is to test whether this morphology is specific to CL-AGN or is also produced by matched non-changing-look samples.

The no-broad-H\(\alpha\) control classes require an [OIII]-based estimate of the optical luminosity.  We therefore place their [OIII]-based bolometric luminosities on the broad-H\(\alpha\) scale using the empirical internal calibration described in Section~\ref{sec:derived} (Equation~\ref{eq:oiii_lbol}).  The calibration changes the median broad-H\(\alpha\)-minus-[OIII] luminosity offset from \(-0.29\) dex to approximately zero, with residual scatter \(0.33\) dex (Appendix~\ref{app:oiii_calibration}).  We use this only as a statistical luminosity scale for the no-broad-H\(\alpha\) controls, not as a physical bolometric correction.

\begin{figure}
    \centering
    \includegraphics[width=\columnwidth]{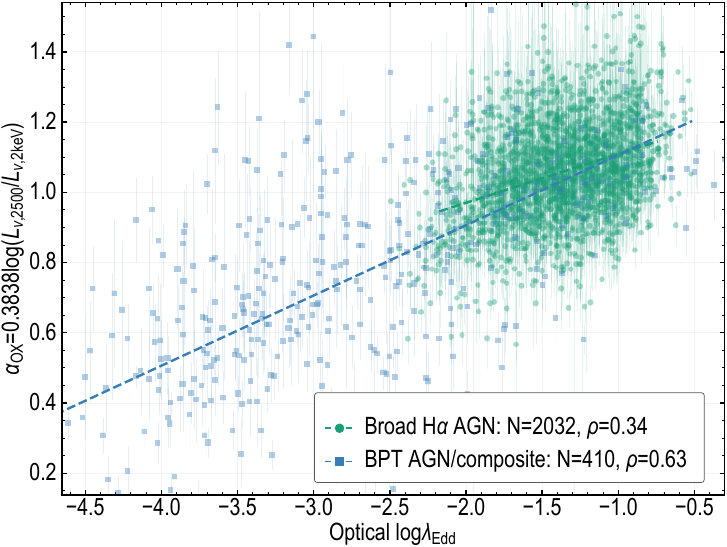}
    \caption{
    \textbf{Apparent \(\alpha_{\rm OX}\)--optical Eddington-ratio relation in matched control samples.}
    Class A broad-H\(\alpha\) AGN and Class B no-broad-H\(\alpha\) BPT AGN/composite galaxies are shown on a common coordinate system, all selected at \(0.35<z<0.45\) with the same 10 arcsec eRASS matching and P3-based \(L_{\rm 2keV}\) measurement.  For Class B, the optical luminosity is placed on the broad-H\(\alpha\) scale using \(L_{\rm bol,[OIII]}=(1.79\times10^3)L_{\rm [OIII]}\) and \(L_{5100}=L_{\rm bol}/9\).  In this redshift window, rest-frame 2 keV lies at observed \(1.38\)--\(1.48\) keV, inside P3.  Dashed lines show separate linear fits for each control class, and faint vertical error bars show the propagated P3-band flux uncertainty in \(\alpha_{\rm OX}\).  The same positive apparent correlation appears in both controls, indicating that the trend is not unique to changing-look systems.
    }
    \label{fig:alphaox_lam_samples}
\end{figure}

For the matched-control test, we impose a common selection on every control class: \(0.35<z<0.45\), eRASS separation \(<10\) arcsec, finite optical \(\lambda_{\rm Edd}\), and a valid P3-based \(L_{\rm 2keV}\).  Figure~\ref{fig:alphaox_lam_samples} shows \(\alpha_{\rm OX}\) as a function of \(\lambda_{\rm Edd,opt}\) for the two non-changing-look controls on a common set of axes: Class A broad-H\(\alpha\) AGN and Class B no-broad-H\(\alpha\) BPT AGN/composite galaxies.  The CL-AGN redshift-window subset is omitted from this comparison figure because the CL-AGN-only view is already shown in Figure~\ref{fig:clagn10_alphaox}, while the purpose here is to test whether the same apparent morphology is present in larger non-changing-look controls.

The same positive apparent relation appears in both non-changing-look control classes.  For Class A, we find \(N=2032\), \(\rho=0.34\), and a fitted slope of \(0.14\).  Class B shows a stronger formal relation, \(N=410\), \(\rho=0.63\), and slope \(0.20\).  The Class B trend should not be interpreted as a clean AGN accretion-rate sequence; its optical continua and [OIII]-based luminosity estimates are more strongly affected by host-galaxy light and by the way the formal optical Eddington ratio is constructed.  Its value here is diagnostic: an \(\alpha_{\rm OX}\)--\(\lambda_{\rm Edd,opt}\) trend can appear even where the physical meaning of \(\lambda_{\rm Edd,opt}\) is uncertain.

The fact that the same apparent relation appears in the control classes is important.  If the \(\alpha_{\rm OX}\)--\(\lambda_{\rm Edd,opt}\) relation were a unique signature of a changing-look accretion-state transition, one would not necessarily expect it to appear with comparable morphology in ordinary broad-line AGN or in host-dominated no-broad-H\(\alpha\) systems.  This behavior is therefore a warning that \(\alpha_{\rm OX}\)--Eddington-ratio morphology alone is not sufficient evidence for a CL-AGN-specific state transition.

\subsection{Black-hole-mass stratification and built-in covariance}
\label{sec:alphaox_covariance}

\begin{figure*}
    \centering
    \includegraphics[width=\textwidth]{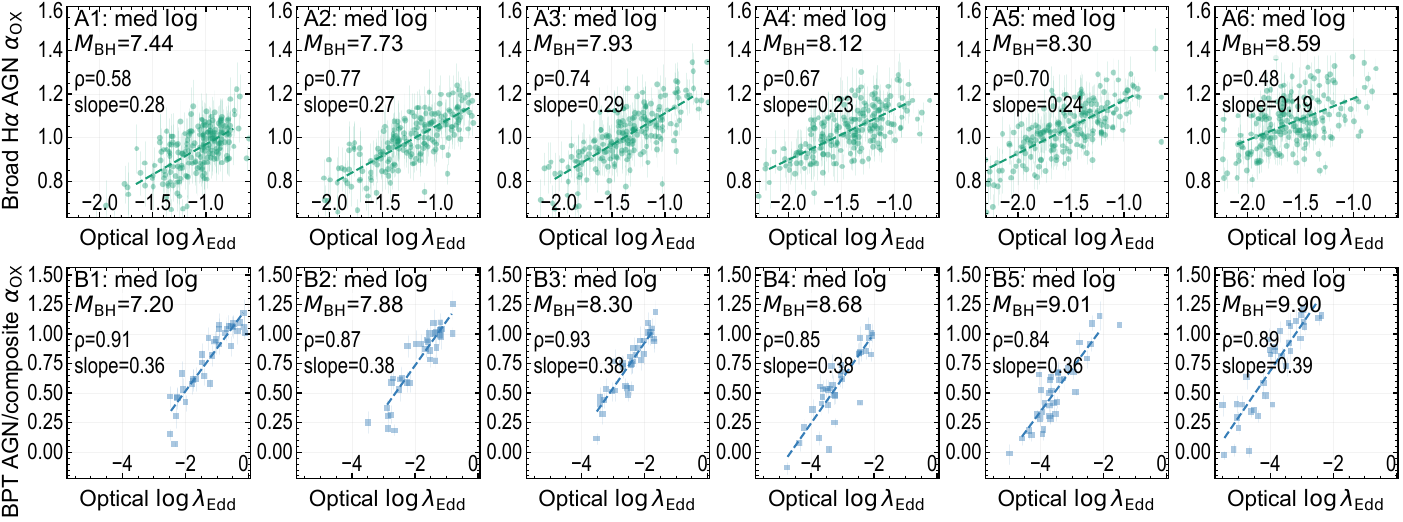}
    \caption{
    \textbf{Black-hole-mass stratification in the control samples.}
    Class A broad-H\(\alpha\) AGN and Class B no-broad-H\(\alpha\) BPT AGN/composite objects with \({\rm DET\_LIKE}_{\rm P3}\ge6\) are divided into six equal-number bins of \(M_{\rm BH}\) within each class.  The two rows are labelled A1--A6 and B1--B6, respectively.  For Class B, \(L_{2500}\) and \(\lambda_{\rm Edd,opt}\) are recomputed from the broad-H\(\alpha\)-calibrated [OIII] luminosity scale, \(L_{\rm bol,[OIII]}=(1.79\times10^3)L_{\rm [OIII]}\).  Each panel corresponds to one \(M_{\rm BH}\) bin and shows the individual sources in that bin.  Dashed lines and panel statistics are computed from a linear fit to the individual points; faint vertical error bars show the propagated P3-band flux uncertainty in \(\alpha_{\rm OX}\).  Class B is a diagnostic control rather than a clean physical accretion sequence.  Its presence reinforces that the apparent structure can be produced by the shared dependence of \(L_{2500}\), \(\lambda_{\rm Edd,opt}\), and \(M_{\rm BH}\).
    }
    \label{fig:alphaox_massbins}
\end{figure*}

Class A is the most suitable population for testing the origin of the apparent relation, because both \(M_{\rm BH}\) and \(\lambda_{\rm Edd,opt}\) are physically interpretable.  Figure~\ref{fig:alphaox_massbins} therefore uses Class A as the primary anchor, but also shows Class B as a diagnostic control.  Within each class, the sources are divided into six bins of black-hole mass.  Each panel shows the individual sources in that mass bin together with a linear fit, so that both the scatter and the fitted trend remain visible.

This behavior follows naturally from the definitions of the quantities.  The optical Eddington ratio is defined from the optical continuum luminosity and black-hole mass,
\begin{equation}
    \lambda_{\rm Edd,opt}
    =
    \frac{k_{5100}L_{5100}}{L_{\rm Edd}},
\label{eq:lambda_opt_projection}
\end{equation}
with \(L_{\rm Edd}\propto M_{\rm BH}\).  Since \(L_{2500}\) is inferred from the same optical continuum, the UV luminosity entering \(\alpha_{\rm OX}\) is not independent of \(\lambda_{\rm Edd,opt}\).  To first order,
\begin{equation}
    \log L_{2500}
    \simeq
    \log \lambda_{\rm Edd,opt}
    +
    \log M_{\rm BH}
    +
    {\rm const.}
    \label{eq:l2500_lambda_mbh}
\end{equation}
Therefore, even if the primary empirical relation is the known \(\alpha_{\rm OX}\)--\(L_{2500}\) dependence \citep{Steffen2006,Just2007,Lusso2010}, it will project into an apparent \(\alpha_{\rm OX}\)--\(\lambda_{\rm Edd,opt}\) relation with a black-hole-mass stratification.

This projection effect is central to the interpretation of Figure~\ref{fig:alphaox_massbins}.  The existence of a positive trend at fixed \(M_{\rm BH}\) does not by itself prove that Eddington ratio is the independent physical driver; within a narrow mass bin, \(\lambda_{\rm Edd,opt}\) is effectively a proxy for the optical/UV luminosity.  Similarly, the shift between mass bins can arise because, at fixed \(\lambda_{\rm Edd,opt}\), higher-mass black holes have higher UV luminosities.  As a stricter X-ray-detection check, Figure~\ref{fig:alphaox_massbins} shows only sources with \({\rm DET\_LIKE}_{\rm P3}\ge6\) and finite derived quantities.  This leaves 1200 Class A sources and 211 Class B sources.  In Class A, all six mass bins retain positive \(\alpha_{\rm OX}\)--\(\lambda_{\rm Edd,opt}\) trends, with fitted slopes of \(0.19\)--\(0.31\).  Class B also shows positive formal relations after applying the broad-H\(\alpha\)-calibrated [OIII] luminosity scale, but these should be interpreted diagnostically: its line-based luminosities and \(M_{\rm BH}\) estimates are more host-dependent.  The same algebraic projection appears even when the physical meaning of the formal Eddington ratio is weak.

\subsection{X-ray-based Eddington ratios and an estimator-induced negative branch}
\label{sec:alphaox_xray_estimator}

The sign of the apparent \(\alpha_{\rm OX}\)--Eddington-ratio relation changes when the Eddington ratio is estimated from the X-ray luminosity.  This is expected from the definitions.  We keep the bolometric correction fixed at \(k_X=20\).  If
\begin{equation}
    \lambda_{\rm Edd,X}
    =
    \frac{k_XL_{2-10}}{L_{\rm Edd}},
\label{eq:lambda_x_projection}
\end{equation}
then \(L_X\) increases the horizontal-axis value.  However, \(L_X\) enters \(\alpha_{\rm OX}\) in the opposite sense: at fixed UV luminosity, a larger X-ray luminosity gives a smaller \(\alpha_{\rm OX}\).  Thus, an X-ray-based Eddington ratio naturally induces a negative branch in the \(\alpha_{\rm OX}\)--Eddington-ratio plane.

\begin{figure*}
    \centering
    \includegraphics[width=\textwidth]{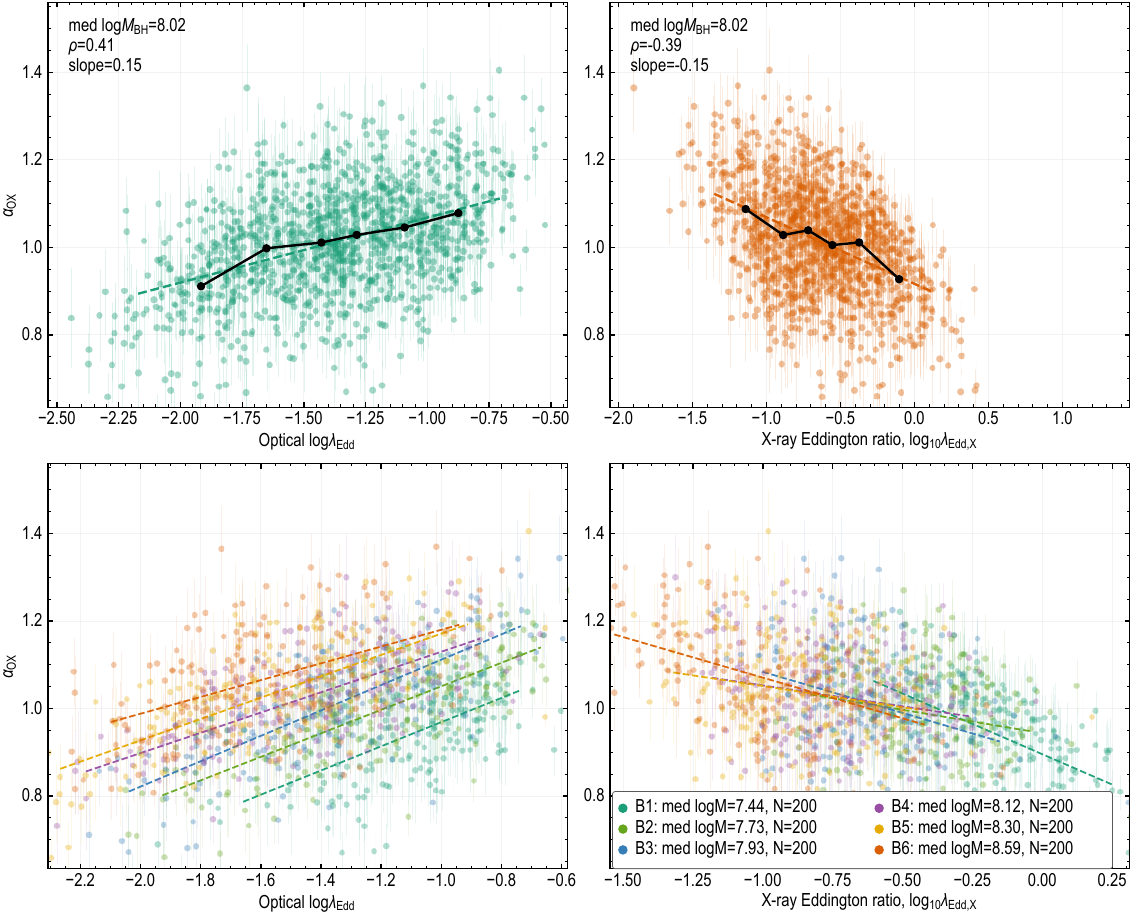}
    \caption{
    \textbf{Estimator dependence of the \(\alpha_{\rm OX}\)--Eddington-ratio plane.}
    Class A broad-H\(\alpha\) AGN with \({\rm DET\_LIKE}_{\rm P3}\ge6\) are shown on optical-Eddington-ratio axes (left column) and X-ray-Eddington-ratio axes (right column).  The top row shows the full sample; black connected points mark binned medians.  The bottom row shows the same sources split into six equal-number bins of \(M_{\rm BH}\), with colours denoting the mass bins and dashed lines showing bin-specific linear fits.  Faint vertical error bars show the propagated P3-band flux uncertainty in \(\alpha_{\rm OX}\).  The optical estimator produces a positive relation because it shares the continuum luminosity used to infer \(L_{2500}\).  The X-ray estimator, here \(k_{\rm bol,X}L_{2-10}/L_{\rm Edd}\), reverses the relation because \(L_X\) enters \(\alpha_{\rm OX}\) and the Eddington-ratio estimator with opposite signs.  A V-shaped morphology can therefore be induced by mixing Eddington-ratio estimators.
    }
    \label{fig:alphaox_estimator}
\end{figure*}

Figure~\ref{fig:alphaox_estimator} demonstrates this estimator dependence for Class A.  When the optical Eddington ratio is used, the sample shows the positive branch discussed above, with \(N=1200\), \(\rho=0.409\), and slope \(0.149\).  When an X-ray-based Eddington ratio is used for the same objects, the relation changes sign, giving \(\rho=-0.394\) and slope \(-0.152\).  The lower panels of Figure~\ref{fig:alphaox_estimator} show the same sources colour-coded by six equal-number black-hole-mass bins.  This makes clear that the optical-axis trend and the X-ray-axis sign change are not driven by a single narrow mass interval.  The sign reversal does not require a change in the physical accretion state; it follows from whether the horizontal axis is built from optical/UV luminosity or X-ray luminosity.  When \(L_{2-10}\) is instead taken from the P1--P5 band-flux fit for Class A, the negative relation is similar or stronger, \(N=1060\), \(\rho=-0.450\), and slope \(-0.143\).

This point is directly relevant to interpretations of a V-shaped \(\alpha_{\rm OX}\)--Eddington-ratio morphology.  If the high-Eddington branch is dominated by optical/UV-based luminosity estimates and the low-Eddington branch is dominated by X-ray-based or \(\alpha_{\rm OX}\)-dependent bolometric corrections, the two branches can have opposite signs by construction.  This test demonstrates that estimator mixing can produce an apparent V-shaped morphology; it does not establish that every reported turnover has this origin.

\subsection{Multivariate regressions}
\label{sec:alphaox_regression}

To quantify which variables drive the observed structure, we compare several regression models for Class A.  We fit \(\alpha_{\rm OX}\) as a function of \(L_{2500}\), \(\lambda_{\rm Edd,opt}\), \(M_{\rm BH}\), and redshift, and compare the residual scatter, Bayesian information criterion \citep[BIC;][]{Schwarz1978}, cross-validated root-mean-square error, and adjusted \(R^2\).  The results are summarized in Table~\ref{tab:alphaox_regression}.

We first assess the practical improvement using residual scatter and adjusted \(R^2\).  The \(L_{2500}\)-only model has a residual scatter of \(0.0870\) and an adjusted \(R^2=0.530\), whereas the \(\lambda_{\rm Edd,opt}\)-only model has a larger scatter of \(0.1142\) and an adjusted \(R^2=0.189\).  Thus, although \(\alpha_{\rm OX}\) and \(\lambda_{\rm Edd,opt}\) are correlated, \(\lambda_{\rm Edd,opt}\) alone is not the primary predictor.

Adding \(M_{\rm BH}\) to either \(L_{2500}\) or \(\lambda_{\rm Edd,opt}\) gives nearly identical model performance.  This is expected from Equation~\ref{eq:l2500_lambda_mbh}: once \(M_{\rm BH}\) is included, \(L_{2500}\) and \(\lambda_{\rm Edd,opt}\) contain almost the same information.  The best-performing model additionally includes redshift, reducing the scatter to \(0.0816\).  Large \(\Delta{\rm BIC}\) values support the model ranking, but can reflect modest improvements in a large sample and should not be interpreted as effect sizes.  We interpret the redshift term primarily as a nuisance term that captures residual luminosity--redshift covariance and eRASS selection effects, rather than as direct evidence for intrinsic redshift evolution.

\begin{table*}
\centering
\caption{Regression models for \(\alpha_{\rm OX}\) in Class A broad-H\(\alpha\) AGN.  The BIC values are reported relative to the best model.}
\label{tab:alphaox_regression}
\begin{tabular}{lcccc}
\hline
Model & Scatter & \(\Delta{\rm BIC}\) & CV RMSE & Adjusted \(R^2\) \\
\hline
\(L_{2500}+M_{\rm BH}+z\) & 0.0816 & 0 & 0.0819 & 0.586 \\
\(L_{2500}+M_{\rm BH}\) & 0.0858 & 114 & 0.0860 & 0.543 \\
\(\lambda_{\rm Edd,opt}+M_{\rm BH}\) & 0.0858 & 114 & 0.0860 & 0.543 \\
\(L_{2500}\) & 0.0870 & 140 & 0.0872 & 0.530 \\
\(\lambda_{\rm Edd,opt}\) & 0.1142 & 795 & 0.1142 & 0.189 \\
\hline
\end{tabular}
\end{table*}

Residual tests lead to the same conclusion.  After subtracting the best-fit \(\alpha_{\rm OX}\)--\(L_{2500}\) relation, the residual dependence on \(M_{\rm BH}\) is weak, with a slope of \(-0.029\) per dex and Spearman coefficient \(\rho=-0.130\).  By contrast, after subtracting the \(\alpha_{\rm OX}\)--\(\lambda_{\rm Edd,opt}\) relation, a strong residual dependence on \(M_{\rm BH}\) remains, with a slope of \(0.165\) per dex and \(\rho=0.575\).  This again indicates that controlling for \(\lambda_{\rm Edd,opt}\) alone does not remove the underlying UV-luminosity dependence.

\subsection{Interim conclusion}
\label{sec:alphaox_summary}

The tests in this section show that the \(\alpha_{\rm OX}\)--Eddington-ratio plane is not a clean standalone diagnostic of disk--corona state transitions.  The apparent positive relation with optical Eddington ratio is present in CL-AGN and in both control classes, and its mass stratification follows naturally from the relation between \(L_{2500}\), \(M_{\rm BH}\), and \(\lambda_{\rm Edd,opt}\).  Using an X-ray-based Eddington ratio induces a negative branch by construction, showing how a V-shaped morphology can be generated by mixing estimators.  Multivariate regressions then confirm that \(L_{2500}\) is a stronger predictor of \(\alpha_{\rm OX}\) than \(\lambda_{\rm Edd,opt}\) alone.  These results do not imply that disk--corona coupling is absent in CL-AGN.  Rather, they show that \(\alpha_{\rm OX}\) alone is insufficient to establish an X-ray-binary-like accretion-state sequence.  In the next section we therefore examine an X-ray spectral-slope diagnostic based on the eRASS effective photon index.

\section{An X-ray Spectral-Slope Diagnostic}
\label{sec:gamma}

The previous section shows that the \(\alpha_{\rm OX}\)--Eddington-ratio plane is strongly affected by the luminosities used to construct the two axes.  This motivates a complementary diagnostic based on the X-ray spectral shape rather than the relative UV-to-X-ray normalization.  In this section we examine the eRASS effective photon index, \(\Gamma_{\rm eff}\), and test whether it depends on accretion rate when the Eddington ratio is estimated independently of the X-ray flux.

\subsection{Motivation}
\label{sec:gamma_motivation}

The X-ray photon index, \(\Gamma\), is a commonly used probe of the coronal spectrum.  In luminous radio-quiet AGN, the hard X-ray photon index has been found to increase with Eddington ratio, usually interpreted as stronger Compton cooling of the corona by disk seed photons at higher accretion rates \citep{Shemmer2006, Shemmer2008, Risaliti2009, Brightman2013, Fanali2013}.  At lower accretion rates, low-luminosity AGN and X-ray binaries can show harder-when-brighter behavior, suggesting a change in the dominant X-ray emission mechanism \citep{GuCao2009, Connolly2016, Yang2015}.  Recent multi-epoch X-ray spectral modeling of changing-state AGN has also suggested that \(\Gamma\) and the soft-excess strength evolve with Eddington ratio \citep{Jana2026}.

The eRASS data used here do not generally provide high-count, source-specific X-ray spectra for the full DESI sample.  We therefore use an effective photon index, \(\Gamma_{\rm eff}\), inferred from eRASS catalog band fluxes \citep{Merloni2024}.  This quantity should not be interpreted as a precise intrinsic rest-frame hard-X-ray photon index.  It is instead a statistical hardness proxy.  The key question is whether this proxy shows a robust trend with an accretion-rate estimator that is independent of the X-ray flux.

\subsection{A naive \texorpdfstring{\(\Gamma_{\rm eff}\)--\(L_X/L_{\rm Edd}\)}{Gammaeff--LX/LEdd} relation}
\label{sec:gamma_xray_artifact}

A superficially similar diagnostic to the \(\Gamma\)--\(L_X/L_{\rm Edd}\) diagram of \citet{Yang2015} would place \(\Gamma_{\rm eff}\) against the X-ray Eddington-scaled luminosity, \(L_{2-10}/L_{\rm Edd}\).  However, in shallow eRASS catalog band-flux data this construction is vulnerable to measurement covariance.  If both \(\Gamma_{\rm eff}\) and \(L_{2-10}\) are inferred from the same band-flux fit, then the fitted spectral slope directly affects the extrapolated hard-band luminosity.  A steeper effective spectrum gives a lower inferred rest-frame \(2\)--\(10\) keV luminosity for a fixed soft-band flux, naturally producing an apparent anti-correlation between \(\Gamma_{\rm eff}\) and \(L_{2-10}/L_{\rm Edd}\).

\begin{figure*}
    \centering
    \includegraphics[width=\textwidth]{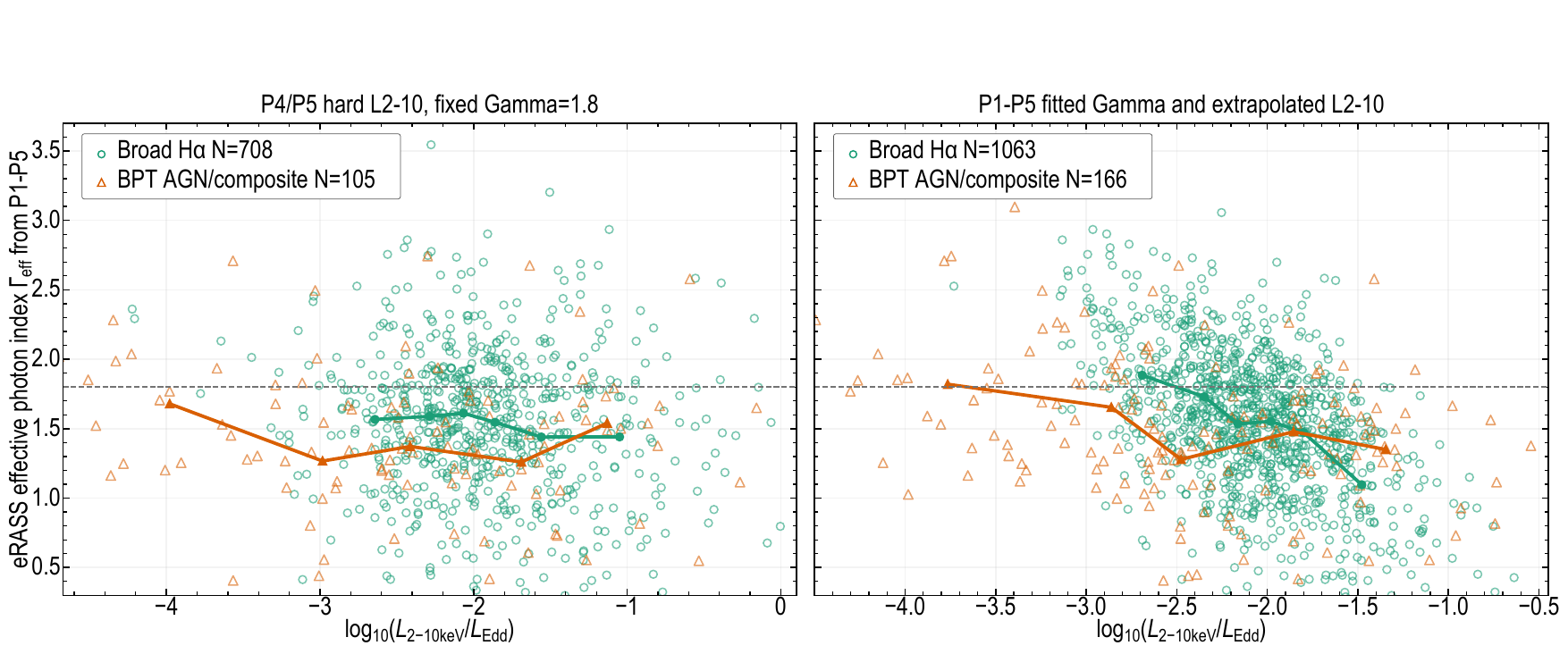}
    \caption{
    \textbf{Systematics of the eRASS \(\Gamma_{\rm eff}\)--\(L_X/L_{\rm Edd}\) diagnostic.}
    The left panel uses a hard-band luminosity estimated from P4/P5 only with fixed \(\Gamma_X=1.8\).  The right panel uses \(L_{2-10}\) extrapolated from the same P1--P5 band-flux fit that yields \(\Gamma_{\rm eff}\).  The strong anti-correlation in the coupled right panel is greatly reduced when the horizontal axis is estimated from hard-band fluxes with fixed \(\Gamma_X\), showing the covariance introduced when the same catalog band-flux fit determines both axes.
    }
    \label{fig:gamma_lx_artifact}
\end{figure*}

Figure~\ref{fig:gamma_lx_artifact} illustrates this effect.  In the full eRASS-detected Class A sample, the coupled implementation in which both \(\Gamma_{\rm eff}\) and \(L_{2-10}\) are derived from the same P1--P5 band-flux fit gives a strong apparent anti-correlation, with Spearman coefficient \(\rho=-0.499\) and a fitted slope of \(-0.685\).
We test this by decoupling the horizontal axis from the fitted \(\Gamma_{\rm eff}\).  When \(L_{2-10}\) is estimated from the non-overlapping hard-band P4/P5 fluxes using a fixed photon index of \(\Gamma_X=1.8\), the correlation weakens substantially, to \(\rho=-0.107\) with a fitted slope of \(-0.117\).
We therefore do not interpret the catalog-band \(\Gamma_{\rm eff}\)--\(L_X/L_{\rm Edd}\) anti-correlation as evidence for a low-accretion state branch.  Instead, we treat it as a measurement-covariance artifact in the eRASS band-flux data.  This statement applies to our eRASS catalog-band effective slopes and should not be read as a critique of high-count X-ray spectral analyses with source-specific response modeling.

\subsection{\texorpdfstring{\(\Gamma_{\rm eff}\)}{Gammaeff} versus optical Eddington ratio}
\label{sec:gamma_optical}

To avoid the direct covariance between \(\Gamma_{\rm eff}\) and \(L_{2-10}\), we next compare \(\Gamma_{\rm eff}\) with the optical Eddington ratio, \(\lambda_{\rm Edd,opt}\).  This horizontal axis is based on the DESI optical continuum and the adopted black-hole mass, and is independent of the eRASS X-ray fluxes.  It therefore provides a cleaner test of whether the effective X-ray spectral slope depends on accretion rate.

\begin{figure*}
    \centering
    \includegraphics[width=\textwidth]{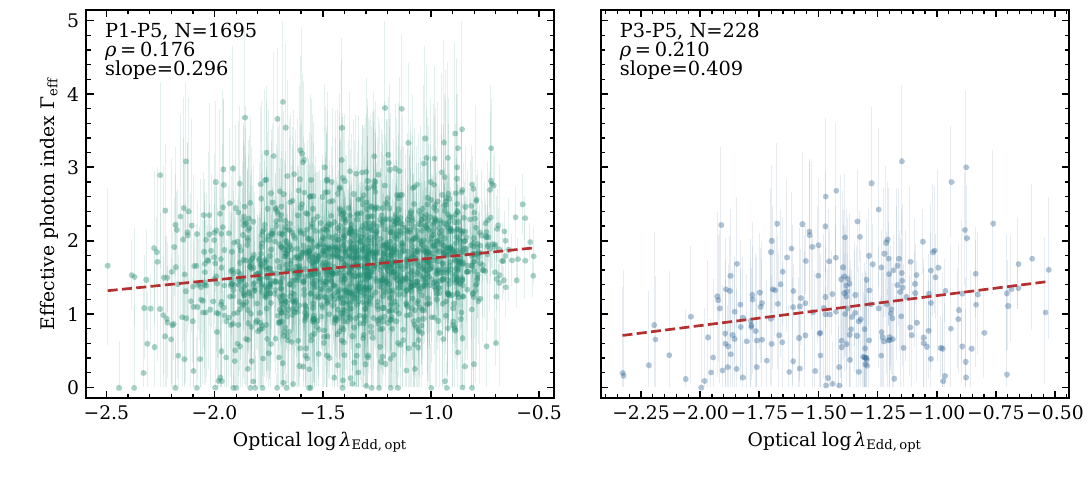}
    \caption{Effective X-ray photon index versus optical Eddington ratio.  Left: the independent P1--P5 Class A sample (1695 objects).  Right: P3--P5 fits for 228 objects after excluding 64 lower-boundary fits from 292 finite results.  Points and vertical bars show the saved best fits and profile \(\Delta\chi^2=1\) intervals; dashed lines are individual-object OLS fits.  The left and right samples differ: the full-band result on the same 228 objects is quoted in the text and Appendix~\ref{app:m2_hard}.  No photon-level or response-aware spectral fit is implied.}
    \label{fig:gamma_lambda}
\end{figure*}

Figure~\ref{fig:gamma_lambda} shows \(\Gamma_{\rm eff}\) versus optical Eddington ratio for the independent Class A sample.  The 1695 P1--P5 fits have \(\rho=0.176\) and slope 0.296.  This is a weak positive catalogue-band association; its physical interpretation requires the same-source hard-band tests below.

The corresponding CL-AGN and no-broad-H\(\alpha\) samples are less constraining for \(\Gamma_{\rm eff}\).  In the \(0.35<z<0.45\) CL-AGN subset, only nine eRASS-matched sources have both a P1--P5 \(\Gamma_{\rm eff}\) estimate and a finite optical Eddington ratio, so they are not used for an independent correlation measurement.  Class B also does not show a meaningful \(\Gamma_{\rm eff}\)--\(\lambda_{\rm Edd,opt}\) sequence: for the independent Class B set, \(N=285\), \(\rho=-0.001\), and slope -0.012.  This reinforces our use of Class A as the cleanest optical-Eddington-ratio anchor for the X-ray spectral-slope test.

The original constant-\(\Gamma_{\rm true}=1.9\) full-band null is conditional on the 1063-object P3-quality subset: its median \(\rho=0.036\), 16th--84th percentiles [0.004,0.066], and 0/1000 exceedances do not establish the significance of the independent 1695-object sample.

P3--P5 alone yields 292 finite fits.  Removing 64 fits at the lower grid boundary \(\Gamma_{\rm eff}=0\) leaves 228 objects.  For these same 228 objects, P1--P5 gives \(\rho=0.433\), slope 0.530, whereas P3--P5 gives \(\rho=0.210\), slope 0.409.  The hard-only paired-object-bootstrap 95 per cent intervals are [0.079,0.337] for \(\rho\) and [0.196,0.611] for slope.  The three hard-band null positive-tail fractions are 0.013, 0.008 and 0.386.  Thus the evidence is conditional on the null model; it is not a uniformly robust hard-band detection.  Weak hard-band information and the post-inspection exclusion of boundary fits prevent attributing the reduced trend uniquely to soft excess.  Full selections, boundary-inclusive comparisons and simulation details are given in Appendix~\ref{app:m2_hard}.

\subsection{Stacked eRASS band fluxes}
\label{sec:gamma_stacking}

As an additional test, we stack the eRASS band fluxes in quartiles of optical Eddington ratio for Class A. This unchanged original stacking display uses the original P3-quality-selected sample, not the independent 1695-object selection.  The purpose of this test is to reduce the impact of individual noisy \(\Gamma_{\rm eff}\) estimates and determine whether the average X-ray spectral hardness changes systematically with \(\lambda_{\rm Edd,opt}\).

The four optical-Eddington-ratio bins have median values \(\log\lambda_{\rm Edd,opt}=-1.818,\allowbreak\ -1.456,\allowbreak\ -1.217,\allowbreak\ -0.912\).
Fitting the stacked band fluxes gives \(\Gamma_{\rm stack}=1.114,\allowbreak\ 1.281,\allowbreak\ 1.328,\allowbreak\ 1.391\), respectively.  Thus, the stacked eRASS spectra show a monotonic softening with increasing optical Eddington ratio.  A linear fit to the stacked values gives a slope of \(0.302\), consistent in sign and approximate magnitude with the individual-source \(\Gamma_{\rm eff}\)--\(\lambda_{\rm Edd,opt}\) relation.

Because this stacking is performed on catalog band fluxes rather than on photon-level spectra with response matrices, we interpret the result as a trend in average effective X-ray hardness rather than as a precision measurement of the intrinsic hard-X-ray photon index.  Nevertheless, the monotonic behavior in the stacked band fluxes provides an independent consistency check that the individual-source trend is not dominated by random band-flux noise.

\begin{figure}
    \centering
    \includegraphics[width=\columnwidth]{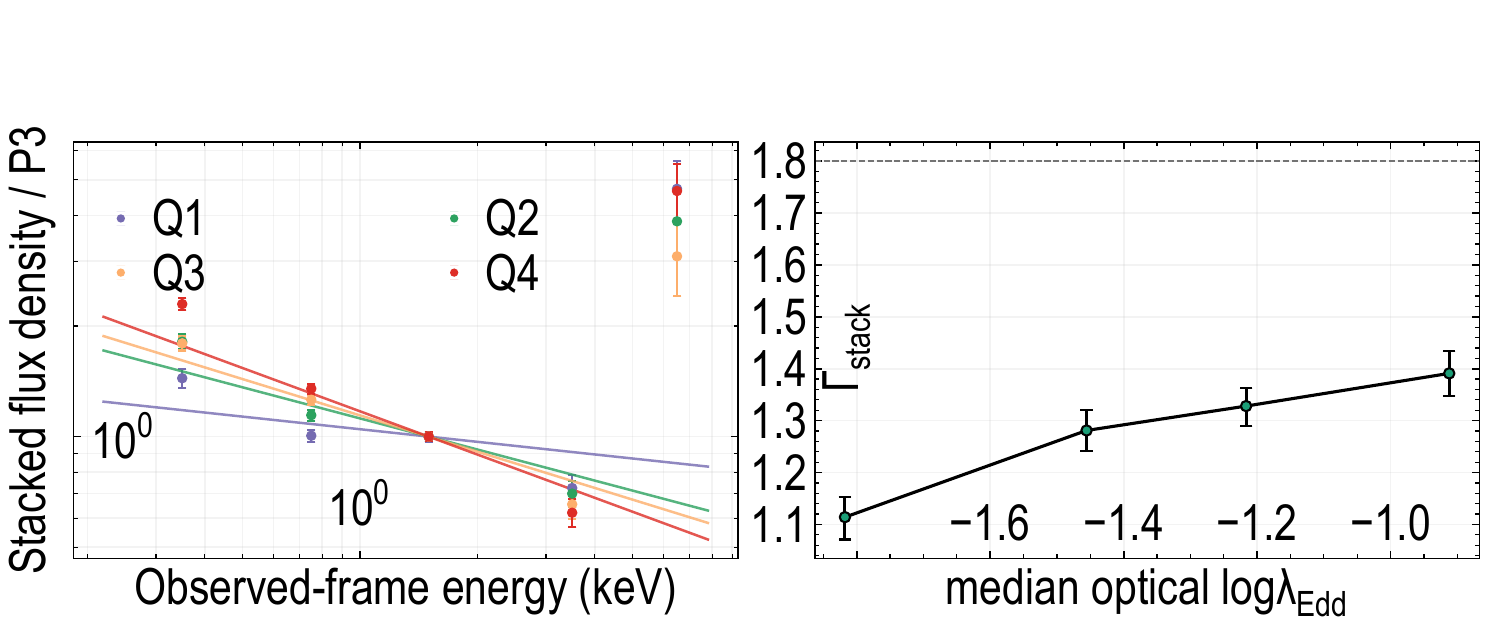}
    \caption{
    \textbf{Stacked eRASS effective photon index in bins of optical Eddington ratio.}
    Class A broad-H\(\alpha\) AGN are divided into quartiles of \(\lambda_{\rm Edd,opt}\).  The stacked eRASS band fluxes show a monotonic increase in \(\Gamma_{\rm stack}\) from the lowest to the highest Eddington-ratio bin, indicating that the average effective X-ray spectrum softens with increasing optical accretion rate.
    }
    \label{fig:gamma_stack}
\end{figure}

\subsection{Interpretation}
\label{sec:gamma_interpretation}

The \(\Gamma_{\rm eff}\)--\(\lambda_{\rm Edd,opt}\) relation provides a more physically interpretable result than the catalog-band \(\Gamma_{\rm eff}\)--\(L_X/L_{\rm Edd}\) diagram.  When the horizontal axis is built from X-ray luminosity inferred from the same band-flux fit, an apparent anti-correlation is expected from covariance between \(\Gamma_{\rm eff}\) and the extrapolated \(L_{2-10}\).  When the horizontal axis instead uses the optical continuum and black-hole mass, Class A shows a weak positive catalogue-band association, but the hard-only tests depend on the null model and boundary selection.

A positive intrinsic relation would be compatible with greater Compton cooling by disk photons, but these model-dependent catalogue-band tests alone do not distinguish that mechanism from spectral-component and selection effects.  Class A primarily probes moderate to high optical Eddington ratios, so the observed trend should be compared to the high-\(\lambda_{\rm Edd}\) branch of previous \(\Gamma\)--Eddington-ratio studies rather than to the low-\(\lambda_{\rm Edd}\) turnover.  We therefore do not claim to detect a full X-ray-binary-like state sequence or a low-accretion-rate reversal in the DESI--eRASS sample.

The main conclusion from this section is methodological as well as physical.  The eRASS effective photon index is not sufficiently precise to replace dedicated XMM-Newton, Chandra, NuSTAR, or Swift/XRT spectral modeling for individual CL-AGN.  Combined with an X-ray-independent optical Eddington ratio, it provides exploratory evidence for a catalogue-band association, whose hard-band significance is conditional on the adopted null model.  This supports the use of X-ray spectral-slope diagnostics, rather than \(\alpha_{\rm OX}\) alone, for future tests of accretion-state changes in CL-AGN.

\section{Discussion}
\label{sec:discussion}

The preceding sections show that the \(\alpha_{\rm OX}\)--Eddington-ratio plane is highly sensitive to how the Eddington ratio is estimated.  The eRASS effective photon index shows a weak positive catalogue-band trend in Class A broad-H\(\alpha\) AGN, but its hard-band interpretation is limited by measurement information and null-model dependence.  In this section we discuss the implications of these results for disk--corona coupling, previous CLQ V-shape claims, and future tests of accretion-state transitions in CL-AGN.

\subsection{What does \texorpdfstring{\(\alpha_{\rm OX}\)}{alphaOX} measure?}
\label{sec:discussion_alphaox}

The optical/UV-to-X-ray slope \(\alpha_{\rm OX}\) is a useful measure of the relative strength of the disk and X-ray continuum components.  It is therefore sensitive to whether an AGN is X-ray weak or X-ray strong relative to its UV emission.  However, it is not a direct measurement of the X-ray spectral state.  It does not measure the shape of the coronal spectrum, the soft-excess strength, the reflection component, or the absorbing column.  This distinction is important for CL-AGN, where changes in the optical/UV luminosity, X-ray luminosity, and spectral components may not occur synchronously.

Our results show that the apparent \(\alpha_{\rm OX}\)--\(\lambda_{\rm Edd,opt}\) relation is not unique to CL-AGN.  Similar positive trends appear in Class A broad-H\(\alpha\) AGN and Class B no-broad-H\(\alpha\) BPT AGN/composite galaxies.  This does not mean that these classes share the same accretion physics.  Rather, it shows that the apparent relation can arise whenever \(L_{2500}\), \(L_{5100}\), and the formal Eddington-ratio estimator are tied to the same optical continuum.

The multivariate regressions reinforce this interpretation.  \(L_{2500}\) is a substantially stronger predictor of \(\alpha_{\rm OX}\) than \(\lambda_{\rm Edd,opt}\) alone, and once \(M_{\rm BH}\) is included, models using \(L_{2500}\) and models using \(\lambda_{\rm Edd,opt}\) become nearly equivalent.  This behavior is expected because \(L_{2500}\), \(\lambda_{\rm Edd,opt}\), and \(M_{\rm BH}\) are algebraically linked.  Thus, the observed black-hole-mass stratification is not necessarily evidence that \(M_{\rm BH}\) independently regulates the disk--corona connection; it can also be the projection of the primary \(\alpha_{\rm OX}\)--UV-luminosity relation.

The X-ray-based Eddington-ratio test provides the complementary point.  When the horizontal axis is built from \(L_X\), the sign of the apparent \(\alpha_{\rm OX}\)--Eddington-ratio relation reverses.  This sign reversal follows directly from the fact that \(L_X\) enters \(\alpha_{\rm OX}\) and \(\lambda_{\rm Edd,X}\) in opposite directions.  Therefore, the same AGN population can show either a positive or a negative branch depending on the adopted Eddington-ratio estimator.  This makes the \(\alpha_{\rm OX}\)--Eddington-ratio plane an estimator-dependent diagram rather than a clean standalone diagnostic of accretion-state transitions.

\subsection{Relation to previous \texorpdfstring{\(\alpha_{\rm OX}\)}{alphaOX} V-shape interpretations}
\label{sec:discussion_vshape}

The idea of an \(\alpha_{\rm OX}\) or X-ray-loudness turnover in AGN was motivated by X-ray-binary scaling arguments.  \citet{Sobolewska2011} predicted that AGN spectral energy distributions should change behavior near a few percent of the Eddington luminosity if they are scaled-up analogues of black-hole X-ray binary state transitions.  This picture motivated later studies of changing-look quasars that placed faded or returning CLQs on the \(\alpha_{\rm OX}\)--Eddington-ratio plane and compared their locations with predicted X-ray-binary-like tracks \citep{Ruan2019, Duffy2025, Gilbert2025}.

These studies provide valuable multiwavelength observations of CLQs, especially because they include X-ray information in states where the optical continuum and broad lines have changed substantially.  However, our results suggest that the interpretation of the \(\alpha_{\rm OX}\)--Eddington-ratio plane requires caution.  The inferred position of a source in this plane depends on how \(L_{\rm bol}\) is estimated.  Optical bolometric corrections can generate a positive branch through the shared optical/UV luminosity.  X-ray bolometric corrections can generate a negative branch through the shared X-ray luminosity.  Bolometric corrections that explicitly depend on the UV-to-X-ray spectral shape introduce still stronger covariance between the two axes \citep{Lusso2010, Lusso2012, Ruan2019}. Although \citet{Ruan2019} recover the \(\alpha_{\rm OX}\) inversion using a UV Eddington ratio without bolometric corrections, UV luminosity still enters both coordinates. A single \(\alpha_{\rm OX}\) value also does not uniquely specify the Eddington ratio along a V-shaped relation, limiting its use as a standalone diagnostic.

This does not imply that previous CLQ observations are uninformative, nor does it rule out accretion-state changes in individual CL-AGN.  Rather, it means that an apparent V-shaped morphology in the \(\alpha_{\rm OX}\)--Eddington-ratio plane is not, by itself, an independent demonstration of an X-ray-binary-like state transition.  A robust test requires Eddington-ratio estimates that are independent of both \(L_X\) and the luminosity used to compute \(L_{2500}\), or else an explicit forward model showing that the observed morphology cannot be produced by the covariance structure of the measured quantities.

The DESI--eRASS control classes provide a direct way to make this point.  If the same \(\alpha_{\rm OX}\)--\(\lambda_{\rm Edd,opt}\) morphology appears in CL-AGN, normal broad-line AGN, and narrow-line AGN/composite systems, then the morphology cannot be attributed uniquely to changing-look accretion-state evolution.  The role of CL-AGN is therefore not simply to populate an existing \(\alpha_{\rm OX}\) diagram, but to test whether the diagram retains physical meaning when compared against matched non-changing-look samples.

\subsection{X-ray spectral slopes as a cleaner state diagnostic}
\label{sec:discussion_gamma}

The eRASS effective photon-index results provide a more constructive route forward.  The catalog-band \(\Gamma_{\rm eff}\)--\(L_X/L_{\rm Edd}\) diagram produces a strong anti-correlation when the same P1--P5 fit supplies both axes.  The trend weakens when the horizontal axis is built from fixed-\(\Gamma\) hard-band luminosities.  This demonstrates that, in shallow eRASS band-flux data, a \(\Gamma_{\rm eff}\)--\(L_X/L_{\rm Edd}\) plane can be dominated by covariance between the fitted slope and the inferred hard-band luminosity.

When the horizontal axis is replaced by the X-ray-independent optical Eddington ratio, Class A shows a positive \(\Gamma_{\rm eff}\)--\(\lambda_{\rm Edd,opt}\) relation.  The conditional full-band null and original quality-selected stacks do not establish robustness for the independent hard-only sample.  Its weaker association can also reflect limited hard-band information, boundary selection and spectral complexity; it cannot be attributed uniquely to soft excess or to Compton cooling \citep{Shemmer2006, Shemmer2008, Risaliti2009, Brightman2013, Fanali2013}.

This result should not be overinterpreted.  Our \(\Gamma_{\rm eff}\) is an effective hardness parameter inferred from eRASS catalog band fluxes, not a true photon index measured from count-level X-ray spectral fitting.  It is therefore not directly equivalent to the \(\Gamma\) values used in detailed XMM-Newton, Chandra, NuSTAR, or Swift/XRT studies.  Moreover, the Class A DESI--eRASS sample primarily probes moderate to high optical Eddington ratios.  It is therefore sensitive to the high-\(\lambda_{\rm Edd}\) coronal-softening branch, but it does not provide a clean test of the low-\(\lambda_{\rm Edd}\) turnover proposed in X-ray-binary-inspired studies \citep{Yang2015, Jana2026}. Related trends occur in general AGN samples: a low-Eddington-ratio anticorrelation \citep{GuCao2009} and opposite branches in \(\Gamma\)--\(L_{2-10}/L_{\rm Edd}\) \citep{Yang2015}. Their strength and universality remain sample-dependent \citep{Trakhtenbrot2017}. \citet{Jana2026} report a V-shaped \(\Gamma\)--Eddington-ratio relation using a common UV-to-X-ray SED-based luminosity prescription. Our sample does not establish a turnover and therefore neither confirms nor rules out such results.

Nevertheless, the contrast between the \(\alpha_{\rm OX}\) and \(\Gamma_{\rm eff}\) results is instructive.  \(\alpha_{\rm OX}\) mixes UV and X-ray luminosities and is strongly affected by how \(L_{\rm bol}\) is estimated.  \(\Gamma_{\rm eff}\), despite its limitations, probes the X-ray spectral shape more directly.  Our results therefore support a practical conclusion: statistical tests of accretion-state transitions in CL-AGN should prioritize X-ray spectral-slope and spectral-component diagnostics over \(\alpha_{\rm OX}\)-based V-shape diagrams.

\subsection{Disk--corona coupling in CL-AGN}
\label{sec:discussion_diskcorona}

The present analysis does not argue against disk--corona coupling in CL-AGN.  On the contrary, many well-studied CL-AGN show strong evidence that the disk and corona evolve together during state changes.  In Mrk~1018, the fading event was accompanied by a strong decrease in the ionizing continuum and soft excess \citep{NodaDone2018, Veronese2024}.  In NGC~1566, the changing-look outburst was associated with the emergence of a Seyfert-like X-ray spectrum \citep{Parker2019, Jana2021}.  In 1ES~1927+654, the disappearance and recovery of the corona demonstrated that the X-ray-emitting region can be dramatically restructured during a changing-look event \citep{Ricci2020, Ricci2021, Masterson2022}.

Our results instead clarify which statistical diagnostics are appropriate for large samples.  A luminosity ratio such as \(\alpha_{\rm OX}\) can identify X-ray-weak or X-ray-bright states relative to the UV, but it does not uniquely determine the physical state of the corona.  The same value of \(\alpha_{\rm OX}\) can be produced by different combinations of disk luminosity, X-ray luminosity, obscuration, spectral slope, and bolometric correction.  By contrast, X-ray spectral diagnostics such as \(\Gamma\), soft-excess strength, reflection fraction, and intrinsic absorption are more directly connected to the structure of the inner accretion flow.

The DESI--eRASS sample occupies an intermediate position between detailed single-object studies and future high-quality X-ray spectral surveys.  It is much larger than typical CL-AGN X-ray monitoring samples, but the eRASS data are too shallow for detailed spectral decomposition of most individual objects.  The value of the present sample is therefore statistical: it can reveal which broad-band diagnostics are robust against covariance and which apparent trends are produced by the way derived quantities are constructed.

\subsection{Limitations}
\label{sec:discussion_limitations}

Several limitations should be kept in mind.  First, the DESI spectra and eRASS X-ray observations are not generally simultaneous.  This prevents us from constructing reliable object-by-object evolutionary tracks in the \(\alpha_{\rm OX}\)--Eddington-ratio plane.  Non-simultaneity adds scatter to the population relations and may be especially important for CL-AGN, whose optical and X-ray states can evolve on timescales of months to years.

Second, the eRASS data provide limited spectral information for most sources.  The effective photon index used here is derived from catalog band fluxes rather than from source-specific photon spectra with response matrices, backgrounds, and detailed spectral models.  The \(\Gamma_{\rm eff}\) trends should therefore be interpreted as trends in average X-ray hardness, not as precision measurements of the intrinsic hard-X-ray photon index.

Third, absorption and soft-excess emission can affect \(\Gamma_{\rm eff}\).  Intrinsic absorption can harden the effective spectrum, while soft excess can steepen it.  If the incidence of absorption or soft excess depends on optical Eddington ratio, then part of the observed \(\Gamma_{\rm eff}\)--\(\lambda_{\rm Edd,opt}\) trend may reflect changes in spectral components rather than a pure change in the coronal power-law slope.  Dedicated X-ray spectral fitting is required to separate these effects.

Fourth, the optical Eddington ratio is physically most meaningful for Class A.  In Class B, the optical continuum can be dominated by the host galaxy, and the adopted black-hole mass and bolometric correction may not trace the instantaneous AGN accretion rate.  We therefore use this class mainly as a control for covariance and selection effects, rather than as a physical accretion-rate sequence.  This caution is supported by our internal mass-estimator checks: [OIII] gas-line widths show large scatter relative to broad-H\(\alpha\) virial masses, while catalogue velocity-dispersion products available for this work lacked the full quality-control information needed for a homogeneous stellar-\(\sigma\) selection.  The pPXF-based no-broad-H\(\alpha\) masses are therefore useful for population-level diagnostics, but Class A remains the cleanest accretion-rate anchor.

Finally, the eRASS detection requirement introduces selection effects.  X-ray-detected samples preferentially include sources that are brighter in the soft X-ray band, less absorbed, or more favorable for detection at the survey depth.  The inclusion of redshift in the \(\alpha_{\rm OX}\) regressions partly captures luminosity--redshift and selection covariance, but a full treatment would require forward modeling of the eRASS selection function and upper limits.

\subsection{Limitations and future low-Eddington-ratio tests}

The present DESI--eRASS analysis is primarily a statistical test of broad-band disk--corona diagnostics. It demonstrates that the apparent $\alpha_{\rm OX}$--Eddington-ratio relation depends strongly on the adopted luminosity and Eddington-ratio estimators, while the eRASS catalogue-band spectral-shape index provides tentative evidence for X-ray spectral softening toward higher optical Eddington ratios in broad-line AGN. However, the current sample does not provide a decisive test of the low-accretion-rate regime predicted by X-ray-binary-inspired accretion-state models. This is because the broad-H$\alpha$ sample is dominated by moderate-to-high optical Eddington ratios, whereas the no-broad-H$\alpha$ control samples rely on more uncertain optical accretion-rate proxies.

The most critical regime is
\begin{equation}
\lambda_{\rm Edd}\lesssim10^{-3},
\label{eq:low_edd_regime}
\end{equation}
where radiatively inefficient accretion flows are expected to emerge and the X-ray spectrum may undergo a transition to a harder-when-brighter branch \citep{NarayanYi1995,YuanNarayan2014}. This is also the regime in which $\alpha_{\rm OX}$ is most susceptible to selection effects, luminosity-estimator covariance, and non-simultaneous optical and X-ray observations, making it difficult to isolate genuine changes in the accretion state.

An equally important question is whether the transition itself occurs at a universal Eddington ratio. Recent work has suggested that the critical accretion rate may depend systematically on black-hole mass rather than being characterized by a single universal threshold \citep{Guo2025ApJ,Zhu2026}. If such a mass dependence exists, the transition luminosity would naturally broaden or shift across AGN populations spanning several orders of magnitude in black-hole mass, potentially contributing to the large scatter observed in current population studies. Establishing whether the critical Eddington ratio is universal, or instead scales with black-hole mass, therefore represents an important next step in understanding accretion-state evolution in supermassive black holes.

To address these questions, we are carrying out follow-up observations of faint CL-AGN and low-Eddington-ratio AGN with \emph{Einstein Probe}. The wide-field soft X-ray monitoring capability of EP, together with rapid follow-up from its focusing X-ray telescope, is well suited to detecting X-ray outbursts, re-brightening events, and low-level variability in sources that are too faint for detailed constraints in the current eRASS survey data \citep{Yuan2022EP,Yuan2025EP}. Such observations will enable us to identify low-accretion-rate AGN entering detectable X-ray states and to obtain time-resolved measurements of their X-ray luminosities and spectral shapes.

A key objective of this programme is to test the low-Eddington-ratio branch using diagnostics that are largely independent of the covariance effects identified in this work. For each EP-detected source, simultaneous or near-simultaneous optical spectroscopy and optical/UV photometry will be required to determine the optical accretion state, estimate an Eddington ratio independently of the X-ray luminosity, and establish whether broad emission lines are present. Combined with response-aware X-ray spectral modelling, these observations will allow direct tests of whether CL-AGN exhibit a genuine reversal in X-ray spectral evolution at low accretion rates.

In this sense, the DESI--eRASS analysis presented here establishes the statistical framework for future studies, while \emph{Einstein Probe} provides the observational capability needed to extend this framework into the low-accretion regime. DESI identifies changing-look and low-accretion candidates, eRASS provides a homogeneous all-sky X-ray baseline, and EP monitoring can probe the faint X-ray states where accretion-state transitions are expected to occur. Ultimately, confirming an X-ray-binary-like accretion-state sequence in AGN will require large homogeneous samples spanning a wide range of black-hole masses, Eddington-ratio estimates independent of X-ray luminosity, and time-resolved X-ray spectroscopy. Such datasets will also provide the first opportunity to determine whether the critical transition Eddington ratio is universal or varies systematically with black-hole mass.

\section{Conclusions}
\label{sec:conclusions}

We have used DESI spectroscopy and eROSITA/eRASS X-ray measurements to test whether the \(\alpha_{\rm OX}\)--Eddington-ratio plane provides an independent diagnostic of disk--corona coupling and accretion-state changes in CL-AGN.  Starting from a parent sample of 561 DESI-selected CL-AGN, we focused the control analysis on sources at \(0.35<z<0.45\), where the eRASS P3 band provides direct leverage on rest-frame 2 keV.  We compared the CL-AGN to two 10 arcsec matched eRASS-detected control classes: Class A broad-H\(\alpha\) AGN and Class B no-broad-H\(\alpha\) BPT AGN/composite galaxies.  Our main conclusions are as follows.

\begin{enumerate}

\item The apparent positive \(\alpha_{\rm OX}\)--\(\lambda_{\rm Edd,opt}\) relation appears in the CL-AGN sample and in all eRASS-matched control classes.  It is therefore not a CL-AGN-specific signature of an accretion-state transition.

\item In Class A, the mass stratification and multivariate regressions show that \(L_{2500}\) is the dominant predictor of \(\alpha_{\rm OX}\).  The apparent \(\lambda_{\rm Edd,opt}\) dependence follows naturally because \(L_{2500}\) is inferred from the optical continuum while \(\lambda_{\rm Edd,opt}\propto L_{5100}/M_{\rm BH}\).

\item When the Eddington ratio is instead built from X-ray luminosity, the apparent \(\alpha_{\rm OX}\)--Eddington-ratio relation changes sign.  An apparent V-shaped morphology can therefore be produced by mixing optical/UV-based and X-ray-based luminosity estimators.

\item The catalog-band \(\Gamma_{\rm eff}\)--\(L_X/L_{\rm Edd}\) anti-correlation is dominated by covariance when the same eRASS P1--P5 band-flux fit supplies both \(\Gamma_{\rm eff}\) and \(L_{2-10}\).  By contrast, \(\Gamma_{\rm eff}\) correlates positively with the X-ray-independent optical Eddington ratio in Class A, but the independent hard-only result depends on the null model, and the original full-band null and stacking results refer to their quality-selected subset.

\item \(\alpha_{\rm OX}\) is useful as a descriptor of relative UV-to-X-ray strength, but it is not a clean standalone state diagnostic.  Future tests of CL-AGN disk--corona state changes should prioritize X-ray spectral slopes and spectral components, combined with Eddington-ratio estimates independent of \(L_X\) and \(\alpha_{\rm OX}\).  The low-\(\lambda_{\rm Edd}\) regime, especially sources accessible to \emph{Einstein Probe} follow-up, remains the key regime for testing a true turnover.

\end{enumerate}

In summary, \(\alpha_{\rm OX}\) is a useful descriptor of relative UV-to-X-ray strength, but it is not a clean standalone accretion-state diagnostic.  Apparent \(\alpha_{\rm OX}\)--Eddington-ratio branches can be induced by the luminosity estimator used to define the Eddington ratio.  X-ray spectral-slope diagnostics, combined with accretion-rate estimates independent of X-ray luminosity and \(\alpha_{\rm OX}\), provide a more robust path for testing disk--corona state changes in CL-AGN.

\section*{Acknowledgements}

 W.J.G. acknowledges financial support from  National Natural Science Foundation of China (NSFC) under grant No. 12503019 and the China Scholarship Council (CSC).

This material is based upon work supported by the U.S. Department of Energy (DOE), Office of Science, Office of High-Energy Physics, under Contract No. DE–AC02–05CH11231, and by the National Energy Research Scientific Computing Center, a DOE Office of Science User Facility under the same contract. Additional support for DESI was provided by the U.S. National Science Foundation (NSF), Division of Astronomical Sciences under Contract No. AST-0950945 to the NSF’s National Optical-Infrared Astronomy Research Laboratory; the Science and Technologies Facilities Council of the United Kingdom; the Gordon and Betty Moore Foundation; the Heising-Simons Foundation; the French Alternative Energies and Atomic Energy Commission (CEA); the National Council of Science and Technology of Mexico (CONACYT); the Ministry of Science and Innovation of Spain (MICINN), and by the DESI Member Institutions: \url{https://www.desi.lbl.gov/collaborating-institutions}.

Any opinions, findings, and conclusions or recommendations expressed in this material are those of the author(s) and do not necessarily reflect the views of the U. S. National Science Foundation, the U. S. Department of Energy, or any of the listed funding agencies.

The authors are honored to be permitted to conduct scientific research on Iolkam Du’ag (Kitt Peak), a mountain with particular significance to the Tohono O’odham Nation.

This work is based on data from eROSITA, the primary instrument aboard SRG, a joint Russian-German science mission supported by the Russian Space Agency (Roscosmos), the Russian Academy of Sciences, represented by its Space Research Institute (IKI), the Deutsches Zentrum für Luftund Raumfahrt (DLR), and the Max Planck Society. This work made use of data from the first eROSITA all-sky survey (eRASS1) data release provided by the German eROSITA consortium.

\section*{Data Availability}

The DESI spectra used in this work are available through the \href{https://data.desi.lbl.gov/doc/releases/dr1/}{DESI DR1 public data release}. The eROSITA/eRASS catalogue quantities used for the X-ray matching and band-flux analysis are available from the \href{https://erosita.mpe.mpg.de/dr1/AllSkySurveyData_dr1/Catalogues_dr1/}{eROSITA-DE DR1 eRASS1 catalogue products}. The derived matched catalogues, saved measurement and uncertainty tables, catalogue-level analysis and plotting code, and frozen figure products are publicly available in the GitHub release \href{https://github.com/guowj-astra/CL-AGN-VII/releases/tag/v1.0.0}{\textit{CL-AGN VII: supporting data and catalogue-level analysis code}}, by Guo W.-J., Zhai S., Yao Z.-H., Cheng H., Hu J., and Jin J.-J. (2026), version~1.0.0 (released 2026 October 7). The release includes file checksums, sample-selection checks and provenance notes that distinguish the retained main-figure products from the aligned appendix sensitivity tests; raw survey spectra are accessed through the public releases above.



\bibliographystyle{mnras}
\bibliography{example} 

\begin{thebibliography}{}
\makeatletter
\relax
\def\mn@urlcharsother{\let\do\@makeother \do\$\do\&\do\#\do\^\do\_\do\%\do\~}
\def\mn@doi{\begingroup\mn@urlcharsother \@ifnextchar [ {\mn@doi@}
  {\mn@doi@[]}}
\def\mn@doi@[#1]#2{\def\@tempa{#1}\ifx\@tempa\@empty \href
  {http://dx.doi.org/#2} {doi:#2}\else \href {http://dx.doi.org/#2} {#1}\fi
  \endgroup}
\def\mn@eprint#1#2{\mn@eprint@#1:#2::\@nil}
\def\mn@eprint@arXiv#1{\href {http://arxiv.org/abs/#1} {{\tt arXiv:#1}}}
\def\mn@eprint@dblp#1{\href {http://dblp.uni-trier.de/rec/bibtex/#1.xml}
  {dblp:#1}}
\def\mn@eprint@#1:#2:#3:#4\@nil{\def\@tempa {#1}\def\@tempb {#2}\def\@tempc
  {#3}\ifx \@tempc \@empty \let \@tempc \@tempb \let \@tempb \@tempa \fi \ifx
  \@tempb \@empty \def\@tempb {arXiv}\fi \@ifundefined
  {mn@eprint@\@tempb}{\@tempb:\@tempc}{\expandafter \expandafter \csname
  mn@eprint@\@tempb\endcsname \expandafter{\@tempc}}}

\bibitem[\protect\citeauthoryear{Abareshi et~al.}{Abareshi
  et~al.}{2022}]{DESIInstrument2022}
Abareshi B.,  et~al., 2022, \mn@doi [AJ] {10.3847/1538-3881/ac882b}, 164, 207

\bibitem[\protect\citeauthoryear{Abdul~Karim et~al.}{Abdul~Karim
  et~al.}{2026}]{DESI2025}
Abdul~Karim M.,  et~al., 2026, \mn@doi [AJ] {10.3847/1538-3881/ae4c43}, 171,
  285

\bibitem[\protect\citeauthoryear{Adame et~al.}{Adame
  et~al.}{2024}]{DESIValidation2024}
Adame A.~G.,  et~al., 2024, \mn@doi [AJ] {10.3847/1538-3881/ad0b08}, 167, 62

\bibitem[\protect\citeauthoryear{Baldwin, Phillips  \& Terlevich}{Baldwin
  et~al.}{1981}]{Baldwin1981}
Baldwin J.~A.,  Phillips M.~M.,   Terlevich R.,  1981, \mn@doi [PASP]
  {10.1086/130766}, 93, 5

\bibitem[\protect\citeauthoryear{Brightman et~al.,}{Brightman
  et~al.}{2013}]{Brightman2013}
Brightman M.,  et~al., 2013, \mn@doi [MNRAS] {10.1093/mnras/stt920}, 433, 2485

\bibitem[\protect\citeauthoryear{Cappellari}{Cappellari}{2017}]{Cappellari2017}
Cappellari M.,  2017, \mn@doi [MNRAS] {10.1093/mnras/stw3020}, 466, 798

\bibitem[\protect\citeauthoryear{Cappellari \& Emsellem}{Cappellari \&
  Emsellem}{2004}]{Cappellari2004}
Cappellari M.,  Emsellem E.,  2004, \mn@doi [PASP] {10.1086/381875}, 116, 138

\bibitem[\protect\citeauthoryear{Connolly, McHardy, Skipper  \&
  Emmanoulopoulos}{Connolly et~al.}{2016}]{Connolly2016}
Connolly S.~D.,  McHardy I.~M.,  Skipper C.~J.,   Emmanoulopoulos D.,  2016,
  \mn@doi [MNRAS] {10.1093/mnras/stw878}, 459, 3963

\bibitem[\protect\citeauthoryear{D{\'i}az et~al.,}{D{\'i}az
  et~al.}{2023}]{Diaz2023}
D{\'i}az Y.,  et~al., 2023, \mn@doi [A\&A] {10.1051/0004-6361/202244678}, 669,
  A114

\bibitem[\protect\citeauthoryear{Done, Gierli{\'n}ski  \& Kubota}{Done
  et~al.}{2007}]{Done2007}
Done C.,  Gierli{\'n}ski M.,   Kubota A.,  2007, \mn@doi [A\&ARv]
  {10.1007/s00159-007-0006-1}, 15, 1

\bibitem[\protect\citeauthoryear{Duffy, Eracleous, Ruan, Yang  \& Runnoe}{Duffy
  et~al.}{2025}]{Duffy2025}
Duffy L.,  Eracleous M.,  Ruan J.~J.,  Yang Q.,   Runnoe J.~C.,  2025, \mn@doi
  [ApJ] {10.3847/1538-4357/adeb73}, 989, 102

\bibitem[\protect\citeauthoryear{Duras et~al.,}{Duras et~al.}{2020}]{Duras2020}
Duras F.,  et~al., 2020, \mn@doi [A\&A] {10.1051/0004-6361/201936817}, 636, A73

\bibitem[\protect\citeauthoryear{Fanali, Caccianiga, Severgnini, Della~Ceca,
  Ballo  \& Carrera}{Fanali et~al.}{2013}]{Fanali2013}
Fanali R.,  Caccianiga A.,  Severgnini P.,  Della~Ceca R.,  Ballo L.,   Carrera
  F.~J.,  2013, \mn@doi [MNRAS] {10.1093/mnras/stt757}, 433, 648

\bibitem[\protect\citeauthoryear{Gilbert et~al.,}{Gilbert
  et~al.}{2025}]{Gilbert2025}
Gilbert O.,  et~al., 2025, \mn@doi [arXiv e-prints]
  {10.48550/arXiv.2508.01933}, p. arXiv:2508.01933

\bibitem[\protect\citeauthoryear{Greene \& Ho}{Greene \&
  Ho}{2005}]{GreeneHo2005}
Greene J.~E.,  Ho L.~C.,  2005, \mn@doi [ApJ] {10.1086/431897}, 630, 122

\bibitem[\protect\citeauthoryear{Greene \& Ho}{Greene \&
  Ho}{2006}]{GreeneHo2006}
Greene J.~E.,  Ho L.~C.,  2006, \mn@doi [ApJ] {10.1086/500353}, 641, 117

\bibitem[\protect\citeauthoryear{Gu \& Cao}{Gu \& Cao}{2009}]{GuCao2009}
Gu M.,  Cao X.,  2009, \mn@doi [MNRAS] {10.1111/j.1365-2966.2009.15277.x}, 399,
  349

\bibitem[\protect\citeauthoryear{{Guo} et~al.,}{{Guo}
  et~al.}{2025a}]{Guo2026CLAGN}
{Guo} W.-J.,  et~al., 2025a, \mn@doi [\apjs] {10.3847/1538-4365/adc124}, \href
  {https://ui.adsabs.harvard.edu/abs/2025ApJS..278...28G} {278, 28}

\bibitem[\protect\citeauthoryear{{Guo} et~al.,}{{Guo}
  et~al.}{2025b}]{Guo2025ApJ}
{Guo} W.-J.,  et~al., 2025b, \mn@doi [\apj] {10.3847/1538-4357/ae1d7b}, \href
  {https://ui.adsabs.harvard.edu/abs/2025ApJ...995..139G} {995, 139}

\bibitem[\protect\citeauthoryear{Gupta et~al.,}{Gupta et~al.}{2024}]{Gupta2024}
Gupta K.~K.,  et~al., 2024, \mn@doi [A\&A] {10.1051/0004-6361/202450567}, 691,
  A203

\bibitem[\protect\citeauthoryear{Heckman, Kauffmann, Brinchmann, Charlot,
  Tremonti  \& White}{Heckman et~al.}{2004}]{Heckman2004}
Heckman T.~M.,  Kauffmann G.,  Brinchmann J.,  Charlot S.,  Tremonti C.,
  White S. D.~M.,  2004, \mn@doi [ApJ] {10.1086/422872}, 613, 109

\bibitem[\protect\citeauthoryear{Ili{\'c}, Poppenhaeger, Dsouza, Wolk,
  Ag{\"u}eros  \& Stelzer}{Ili{\'c} et~al.}{2023}]{Ilic2023eRASSMdwarfs}
Ili{\'c} N.,  Poppenhaeger K.,  Dsouza D.,  Wolk S.~J.,  Ag{\"u}eros M.~A.,
  Stelzer B.,  2023, \mn@doi [MNRAS] {10.1093/mnras/stad2277}, 524, 5954

\bibitem[\protect\citeauthoryear{Jana, Kumari, Nandi, Naik, Chatterjee,
  Jaisawal, Hayasaki  \& Ricci}{Jana et~al.}{2021}]{Jana2021}
Jana A.,  Kumari N.,  Nandi P.,  Naik S.,  Chatterjee A.,  Jaisawal G.~K.,
  Hayasaki K.,   Ricci C.,  2021, \mn@doi [MNRAS] {10.1093/mnras/stab2155},
  507, 687

\bibitem[\protect\citeauthoryear{Jana et~al.,}{Jana et~al.}{2023}]{Jana2023}
Jana A.,  et~al., 2023, \mn@doi [MNRAS] {10.1093/mnras/stad2140}, 524, 4670

\bibitem[\protect\citeauthoryear{Jana et~al.,}{Jana et~al.}{2025}]{Jana2025}
Jana A.,  et~al., 2025, \mn@doi [A\&A] {10.1051/0004-6361/202451058}, 693, A35

\bibitem[\protect\citeauthoryear{Jana et~al.,}{Jana et~al.}{2026}]{Jana2026}
Jana A.,  et~al., 2026, \mn@doi [A\&A] {10.1051/0004-6361/202556654}, 707, A213

\bibitem[\protect\citeauthoryear{Just, Brandt, Shemmer, Steffen, Schneider,
  Chartas  \& Garmire}{Just et~al.}{2007}]{Just2007}
Just D.~W.,  Brandt W.~N.,  Shemmer O.,  Steffen A.~T.,  Schneider D.~P.,
  Chartas G.,   Garmire G.~P.,  2007, \mn@doi [ApJ] {10.1086/519990}, 665, 1004

\bibitem[\protect\citeauthoryear{Kaspi, Smith, Netzer, Maoz, Jannuzi  \&
  Giveon}{Kaspi et~al.}{2000}]{Kaspi2000}
Kaspi S.,  Smith P.~S.,  Netzer H.,  Maoz D.,  Jannuzi B.~T.,   Giveon U.,
  2000, \mn@doi [ApJ] {10.1086/308704}, 533, 631

\bibitem[\protect\citeauthoryear{Kauffmann et~al.,}{Kauffmann
  et~al.}{2003}]{Kauffmann2003}
Kauffmann G.,  et~al., 2003, \mn@doi [MNRAS]
  {10.1046/j.1365-2966.2003.07154.x}, 346, 1055

\bibitem[\protect\citeauthoryear{Kewley, Dopita, Sutherland, Heisler  \&
  Trevena}{Kewley et~al.}{2001}]{Kewley2001}
Kewley L.~J.,  Dopita M.~A.,  Sutherland R.~S.,  Heisler C.~A.,   Trevena J.,
  2001, \mn@doi [ApJ] {10.1086/321545}, 556, 121

\bibitem[\protect\citeauthoryear{Kormendy \& Ho}{Kormendy \&
  Ho}{2013}]{KormendyHo2013}
Kormendy J.,  Ho L.~C.,  2013, \mn@doi [ARA\&A]
  {10.1146/annurev-astro-082708-101811}, 51, 511

\bibitem[\protect\citeauthoryear{LaMassa et~al.,}{LaMassa
  et~al.}{2015}]{LaMassa2015}
LaMassa S.~M.,  et~al., 2015, \mn@doi [ApJ] {10.1088/0004-637X/800/2/144}, 800,
  144

\bibitem[\protect\citeauthoryear{Liu, Gu, Lu, Liu  \& Liu}{Liu
  et~al.}{2025}]{Liu2025eRASSGaia}
Liu H.-B.,  Gu W.-M.,  Lu Y.,  Liu T.,   Liu J.-Z.,  2025, \mn@doi [ApJ]
  {10.3847/1538-4357/adfecd}, 991, 125

\bibitem[\protect\citeauthoryear{L{\'o}pez et~al.,}{L{\'o}pez
  et~al.}{2024}]{Lopez2024}
L{\'o}pez I.~E.,  et~al., 2024, \mn@doi [A\&A] {10.1051/0004-6361/202450510},
  692, A209

\bibitem[\protect\citeauthoryear{Lusso et~al.,}{Lusso et~al.}{2010}]{Lusso2010}
Lusso E.,  et~al., 2010, \mn@doi [A\&A] {10.1051/0004-6361/200913298}, 512, A34

\bibitem[\protect\citeauthoryear{Lusso et~al.,}{Lusso et~al.}{2012}]{Lusso2012}
Lusso E.,  et~al., 2012, \mn@doi [MNRAS] {10.1111/j.1365-2966.2012.21513.x},
  425, 623

\bibitem[\protect\citeauthoryear{Lyu, Yan, Wu, Liu  et~al.}{Lyu
  et~al.}{2025}]{Lyu2025}
Lyu B.,  Yan Z.,  Wu X.-B.,  Liu H.,   et~al., 2025, \mn@doi [MNRAS]
  {10.1093/mnras/staf109}, 537, 1099

\bibitem[\protect\citeauthoryear{MacLeod et~al.,}{MacLeod
  et~al.}{2019}]{MacLeod2019}
MacLeod C.~L.,  et~al., 2019, \mn@doi [ApJ] {10.3847/1538-4357/ab05e2}, 874, 8

\bibitem[\protect\citeauthoryear{Masterson et~al.,}{Masterson
  et~al.}{2022}]{Masterson2022}
Masterson M.,  et~al., 2022, \mn@doi [ApJ] {10.3847/1538-4357/ac76c0}, 934, 35

\bibitem[\protect\citeauthoryear{Merloni et~al.}{Merloni
  et~al.}{2024}]{Merloni2024}
Merloni A.,  et~al., 2024, \mn@doi [A\&A] {10.1051/0004-6361/202347165}, 682,
  A34

\bibitem[\protect\citeauthoryear{Narayan \& Yi}{Narayan \&
  Yi}{1995}]{NarayanYi1995}
Narayan R.,  Yi I.,  1995, \mn@doi [ApJ] {10.1086/176343}, 452, 710

\bibitem[\protect\citeauthoryear{Noda \& Done}{Noda \&
  Done}{2018}]{NodaDone2018}
Noda H.,  Done C.,  2018, \mn@doi [MNRAS] {10.1093/mnras/sty2032}, 480, 3898

\bibitem[\protect\citeauthoryear{Parker et~al.,}{Parker
  et~al.}{2019}]{Parker2019}
Parker M.~L.,  et~al., 2019, \mn@doi [MNRAS] {10.1093/mnrasl/sly224}, 483, L88

\bibitem[\protect\citeauthoryear{Predehl et~al.}{Predehl
  et~al.}{2021}]{Predehl2021}
Predehl P.,  et~al., 2021, \mn@doi [A\&A] {10.1051/0004-6361/202039313}, 647,
  A1

\bibitem[\protect\citeauthoryear{Pucha et~al.,}{Pucha et~al.}{2025}]{Pucha2025}
Pucha R.,  et~al., 2025, ApJ, 982, 10

\bibitem[\protect\citeauthoryear{Pucha et~al.,}{Pucha et~al.}{2026}]{Pucha2026}
Pucha R.,  et~al., 2026, arXiv e-prints, p. arXiv:2606.02699

\bibitem[\protect\citeauthoryear{Remillard \& McClintock}{Remillard \&
  McClintock}{2006}]{RemillardMcClintock2006}
Remillard R.~A.,  McClintock J.~E.,  2006, \mn@doi [ARA\&A]
  {10.1146/annurev.astro.44.051905.092532}, 44, 49

\bibitem[\protect\citeauthoryear{Ricci \& Trakhtenbrot}{Ricci \&
  Trakhtenbrot}{2023}]{RicciTrakhtenbrot2023}
Ricci C.,  Trakhtenbrot B.,  2023, \mn@doi [NatAs]
  {10.1038/s41550-023-02108-4}, 7, 1282

\bibitem[\protect\citeauthoryear{Ricci et~al.,}{Ricci et~al.}{2020}]{Ricci2020}
Ricci C.,  et~al., 2020, \mn@doi [ApJL] {10.3847/2041-8213/ab91a1}, 898, L1

\bibitem[\protect\citeauthoryear{Ricci et~al.,}{Ricci et~al.}{2021}]{Ricci2021}
Ricci C.,  et~al., 2021, \mn@doi [ApJS] {10.3847/1538-4365/abe94b}, 255, 7

\bibitem[\protect\citeauthoryear{Richards et~al.,}{Richards
  et~al.}{2006}]{Richards2006}
Richards G.~T.,  et~al., 2006, \mn@doi [ApJS] {10.1086/506525}, 166, 470

\bibitem[\protect\citeauthoryear{Risaliti, Young  \& Elvis}{Risaliti
  et~al.}{2009}]{Risaliti2009}
Risaliti G.,  Young M.,   Elvis M.,  2009, \mn@doi [ApJL]
  {10.1088/0004-637X/700/1/L6}, 700, L6

\bibitem[\protect\citeauthoryear{Ruan, Anderson, Eracleous, Green, Haggard,
  MacLeod, Runnoe  \& Sobolewska}{Ruan et~al.}{2019}]{Ruan2019}
Ruan J.~J.,  Anderson S.~F.,  Eracleous M.,  Green P.~J.,  Haggard D.,  MacLeod
  C.~L.,  Runnoe J.~C.,   Sobolewska M.~A.,  2019, \mn@doi [ApJ]
  {10.3847/1538-4357/ab3c1a}, 883, 76

\bibitem[\protect\citeauthoryear{Schwarz}{Schwarz}{1978}]{Schwarz1978}
Schwarz G.,  1978, \mn@doi [Annals of Statistics] {10.1214/aos/1176344136}, 6,
  461

\bibitem[\protect\citeauthoryear{Shemmer, Brandt, Netzer, Maiolino  \&
  Kaspi}{Shemmer et~al.}{2006}]{Shemmer2006}
Shemmer O.,  Brandt W.~N.,  Netzer H.,  Maiolino R.,   Kaspi S.,  2006, \mn@doi
  [ApJL] {10.1086/506911}, 646, L29

\bibitem[\protect\citeauthoryear{Shemmer, Brandt, Netzer, Maiolino  \&
  Kaspi}{Shemmer et~al.}{2008}]{Shemmer2008}
Shemmer O.,  Brandt W.~N.,  Netzer H.,  Maiolino R.,   Kaspi S.,  2008, \mn@doi
  [ApJ] {10.1086/588776}, 682, 81

\bibitem[\protect\citeauthoryear{Sobolewska, Siemiginowska  \&
  Gierli{\'n}ski}{Sobolewska et~al.}{2011}]{Sobolewska2011}
Sobolewska M.~A.,  Siemiginowska A.,   Gierli{\'n}ski M.,  2011, \mn@doi
  [MNRAS] {10.1111/j.1365-2966.2011.18295.x}, 413, 2259

\bibitem[\protect\citeauthoryear{Steffen, Strateva, Brandt, Alexander,
  Koekemoer, Lehmer, Schneider  \& Vignali}{Steffen et~al.}{2006}]{Steffen2006}
Steffen A.~T.,  Strateva I.~V.,  Brandt W.~N.,  Alexander D.~M.,  Koekemoer
  A.~M.,  Lehmer B.~D.,  Schneider D.~P.,   Vignali C.,  2006, \mn@doi [AJ]
  {10.1086/503627}, 131, 2826

\bibitem[\protect\citeauthoryear{Sunyaev et~al.}{Sunyaev
  et~al.}{2021}]{Sunyaev2021}
Sunyaev R.,  et~al., 2021, \mn@doi [A\&A] {10.1051/0004-6361/202141179}, 656,
  A132

\bibitem[\protect\citeauthoryear{Trakhtenbrot et~al.,}{Trakhtenbrot
  et~al.}{2017}]{Trakhtenbrot2017}
Trakhtenbrot B.,  et~al., 2017, \mn@doi [MNRAS] {10.1093/mnras/stx1117}, 470,
  800

\bibitem[\protect\citeauthoryear{Trakhtenbrot et~al.,}{Trakhtenbrot
  et~al.}{2019}]{Trakhtenbrot2019}
Trakhtenbrot B.,  et~al., 2019, \mn@doi [ApJ] {10.3847/1538-4357/ab39e4}, 883,
  94

\bibitem[\protect\citeauthoryear{Vanden~Berk et~al.}{Vanden~Berk
  et~al.}{2001}]{VandenBerk2001}
Vanden~Berk D.~E.,  et~al., 2001, \mn@doi [AJ] {10.1086/321167}, 122, 549

\bibitem[\protect\citeauthoryear{Veronese, Vignali, Severgnini, Matzeu  \&
  Cignoni}{Veronese et~al.}{2024}]{Veronese2024}
Veronese S.,  Vignali C.,  Severgnini P.,  Matzeu G.~A.,   Cignoni M.,  2024,
  \mn@doi [A\&A] {10.1051/0004-6361/202348098}, 683, A131

\bibitem[\protect\citeauthoryear{Vestergaard \& Peterson}{Vestergaard \&
  Peterson}{2006}]{VestergaardPeterson2006}
Vestergaard M.,  Peterson B.~M.,  2006, \mn@doi [ApJ] {10.1086/500572}, 641,
  689

\bibitem[\protect\citeauthoryear{Yang, Xie, Yuan, Zdziarski, Gierli{\'n}ski, Ho
   \& Yu}{Yang et~al.}{2015}]{Yang2015}
Yang Q.-X.,  Xie F.-G.,  Yuan F.,  Zdziarski A.~A.,  Gierli{\'n}ski M.,  Ho
  L.~C.,   Yu Z.,  2015, \mn@doi [MNRAS] {10.1093/mnras/stu2571}, 447, 1692

\bibitem[\protect\citeauthoryear{Yang et~al.,}{Yang et~al.}{2023}]{Yang2023}
Yang Q.,  et~al., 2023, \mn@doi [ApJ] {10.3847/1538-4357/acdedd}, 953, 61

\bibitem[\protect\citeauthoryear{Yuan \& Narayan}{Yuan \&
  Narayan}{2014}]{YuanNarayan2014}
Yuan F.,  Narayan R.,  2014, \mn@doi [ARA\&A]
  {10.1146/annurev-astro-082812-141003}, 52, 529

\bibitem[\protect\citeauthoryear{Yuan, Zhang, Chen  \& Ling}{Yuan
  et~al.}{2022}]{Yuan2022EP}
Yuan W.,  Zhang C.,  Chen Y.,   Ling Z.,  2022, \mn@doi [arXiv e-prints]
  {10.48550/arXiv.2209.09763}, p. arXiv:2209.09763

\bibitem[\protect\citeauthoryear{Yuan et~al.,}{Yuan et~al.}{2025}]{Yuan2025EP}
Yuan W.,  et~al., 2025, \mn@doi [arXiv e-prints] {10.48550/arXiv.2501.07362},
  p. arXiv:2501.07362

\bibitem[\protect\citeauthoryear{{Zhu}, {Wang}, {Gupta}, {Zeng}, {Devanand},
  {Dogra}, {Bhowmick}  \& {Zhang}}{{Zhu} et~al.}{2026}]{Zhu2026}
{Zhu} L.,  {Wang} Z.,  {Gupta} A.~C.,  {Zeng} Y.,  {Devanand} P.~U.,  {Dogra}
  K.,  {Bhowmick} R.,   {Zhang} J.-J.,  2026, \mn@doi [arXiv e-prints]
  {10.48550/arXiv.2607.15564}, \href
  {https://ui.adsabs.harvard.edu/abs/2026arXiv260715564Z} {p. arXiv:2607.15564}

\makeatother
\end{thebibliography}




\appendix

\section{Internal [OIII] luminosity-scale calibration}
\label{app:oiii_calibration}

The no-broad-H\(\alpha\) control samples require a narrow-line luminosity scale.  We use the broad-H\(\alpha\) calibration sample only to define an internal empirical normalization for this work.

\begin{figure}
    \centering
    \includegraphics[width=\columnwidth]{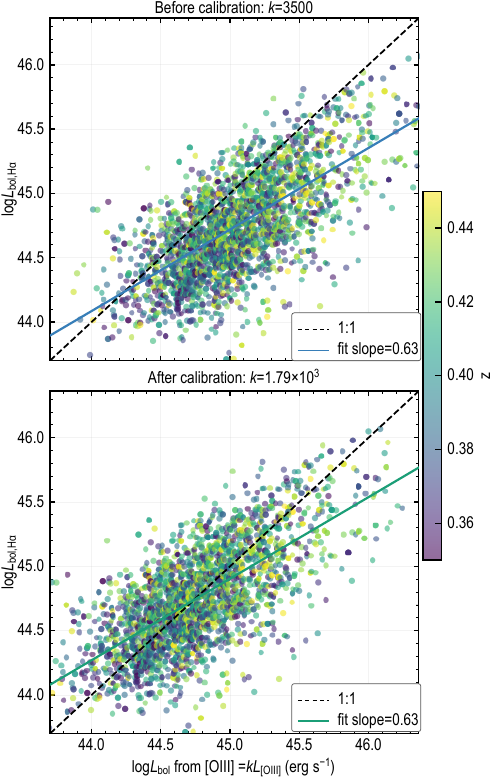}
    \caption{
    \textbf{Internal broad-H\(\alpha\) calibration of [OIII]-based bolometric luminosity.}
    For the \(0.35<z<0.45\) DESI QSO broad-H\(\alpha\) calibration sample, the broad-H\(\alpha\)-based bolometric luminosity, \(L_{\rm bol}=9L_{5100}\), is compared with [OIII]-based luminosities.  The upper panel uses the conventional uncalibrated \(L_{\rm bol,[OIII]}=3500L_{\rm [OIII]}\) conversion.  The lower panel uses the internal empirical normalization \(k_{\rm [OIII]}=1.79\times10^3\).  The dashed line is the one-to-one relation, solid lines are linear fits, and points are coloured by redshift.  For the calibration sample (\(N=2687\)), the median \(L_{\rm bol,H\alpha}\)-minus-[OIII] offset changes from \(-0.29\) dex before calibration to approximately zero after calibration, with residual scatter \(0.33\) dex and Pearson \(r=0.68\).  This normalization is used only to place the no-broad-H\(\alpha\) controls on the broad-H\(\alpha\) luminosity scale; it is not a universal physical bolometric correction.
    }
    \label{fig:oiii_lbol_calibration}
\end{figure}



\begingroup
The constant normalization corrects the zero point only: the calibration slope remains 0.634, rather than unity.  As an internal sensitivity test, we use the same 2687 calibrators and compare the fixed coefficient with the complete ordinary-least-squares relation
\begin{equation}
\log\!\left(\frac{L_{\rm bol}}{\mathrm{erg\,s^{-1}}}\right)
=45.0644+0.6341\left[\log\!\left(\frac{L_{\rm [OIII]}}{\mathrm{erg\,s^{-1}}}\right)-42\right].
\label{eq:m4_powerlaw}
\end{equation}
The unrounded fixed coefficient is 1785.33993698; it is displayed as \(1.79\times10^3\), not as a precisely known physical correction.  The calibration residual standard deviation is 0.330 dex for the fixed coefficient and 0.290 dex for the power law.  Separate 5000-object-bootstrap calibration intervals (95 per cent) are [45.0467,45.0817] for the pivot intercept and [0.6074,0.6610] for the slope.

For the same 410 primary Class B sources, we recalculate both optical quantities under each calibration: \(L_{5100}=L_{\rm bol}/9\), then the same optical-to-2500-\AA\ extrapolation and \(\lambda_{\rm Edd,opt}=L_{\rm bol}/L_{\rm Edd}\).  Masses and P3 X-ray luminosities are fixed.  Combining Equations~\ref{eq:alphaox} and~\ref{eq:lambda_opt} gives
\begin{equation}
\Delta\alpha_{\rm OX}=0.3838\,\Delta\log L_{\rm bol}
=0.3838\,\Delta\log\lambda_{\rm Edd,opt}.
\label{eq:m4_shared_shift}
\end{equation}
The coordinates therefore move together; this is not an independent luminosity validation.  The full Class B correlation changes from 0.636 to 0.458 and the slope from 0.200 to 0.131.  These are formal narrow-line-based diagnostic relations, not a physical accretion sequence.

The calibrators span \(39.9129\le\log[L_{\rm [OIII]}/(\mathrm{erg\,s^{-1}})]\le43.2811\).  Of the 410 objects, 344 lie inside this domain and 66 below it; none lies above.  Of the 66 extrapolators, 65 have [OIII] flux/error below 3.  The largest individual optical-scale shift is 2.72 dex, so low-luminosity extrapolation is not reliable as an individual AGN luminosity measurement.  For the same 344 in-domain objects, the correlation decreases from 0.551 to 0.367 and the slope from 0.142 to 0.085.
The 95 per cent intervals for the power-law-minus-fixed slope change are \([-0.079,-0.060]\) for all 410 objects and \([-0.065,-0.051]\) in-domain.  These intervals use 5000 paired object resamples.

The paired intervals condition on fixed calibration coefficients; the separate calibration bootstrap, measurement uncertainties and individual predictive scatter are not propagated into them.  Seventeen TARGETIDs overlap the calibrator and Class B sets, so these samples cannot be assumed independent.  The sensitivity demonstrates calibration-dependent trend strength; neither relation constitutes an external physical calibration.  No Class A quantities, main-figure calibrations or other diagnostic classes are changed by this Appendix comparison.
\endgroup

\begingroup
\section{Band and bolometric-correction sensitivity tests}
\subsection{P4-anchored rest-frame 2 keV extrapolation}
\label{app:m1_p4}
This test uses the 2032-object Class A primary sample: 10 arcsec matching, positive P3 flux and finite optical quantities, with no P3 likelihood threshold.  The separately defined P3-quality sample has 1200 objects.  We hold the adopted optical quantities fixed and normalize a power law of \(\Gamma_X=1.8\) to observed P4 (2--5 keV), extrapolating to \(E_0=2/(1+z)\) keV:
\begin{equation}
L_{\nu,\mathrm{P4}}(2\,\mathrm{keV})=
\frac{4\pi D_L^2h}{1+z}
\frac{F_{\mathrm{P4}}E_0^{1-\Gamma_X}}
{\int_2^5 E^{1-\Gamma_X}\,\mathrm{d}E}.
\label{eq:p4_lnu_2kev}
\end{equation}
Here \(h\) is in keV s and energies are in observed keV.  We recompute \(\alpha_{\rm OX}=0.3838\log(L_{2500}/L_\nu)\) without changing the optical horizontal axis.  P4 does not directly contain the redshifted 2 keV point in this window.  The calculation neither measures nor removes a soft-excess component and is not a response-aware spectral fit.

Positive finite P3/P4 fluxes and errors, finite P4 likelihood and required inputs leave 1053 paired sources with no additional P4 likelihood cut, 314 at \({\rm DET\_LIKE}_{\rm P4}\ge3\), and 127 at \(\ge6\).  We compare P3 and P4 on identical objects at each threshold, rather than comparing a hard-detected subset to the whole soft sample.
The paired P3/P4 slopes are 0.135/0.172 without a P4 likelihood cut, 0.172/0.219 at likelihood 3, and 0.118/0.213 at likelihood 6.  The corresponding 95 per cent intervals for the P4-minus-P3 slope change are \([-0.012,0.090]\), \([-0.009,0.105]\), and \([0.013,0.189]\), respectively.

At each threshold, 5000 paired source resamples keep optical quantities fixed and perturb the P3 and P4 fluxes independently with mean-preserving lognormal distributions.  For a catalogue flux \(F\) and error \(s_F\), we use \(s_{\ln F}^2=\ln[1+(s_F/F)^2]\) and \(\ln F'=\ln F-s_{\ln F}^2/2+s_{\ln F}\epsilon\), with standard-normal \(\epsilon\).  These are conditional perturbation percentile intervals, not an intrinsic-scatter correction; cross-band error covariance and selection uncertainty are unavailable.  The first two slope-difference intervals include zero, whereas the likelihood-6 interval does not; significance is conditional on the stated perturbation model.

The median P4 flux/error ratios are 0.95, 1.53 and 1.87; even at likelihood 6, 68.5 per cent have flux/error below 2.  The likelihood cut is therefore not interchangeable with a Gaussian flux-SNR threshold.  Hard-band selection, noisy extrapolation and unknown spectral complexity limit this test.  Agreement cannot exclude soft excess, and a difference could not uniquely identify it.

\subsection{P3--P5 effective-index sensitivity}
\label{app:m2_hard}
The independent 1695-object Class A parent requires at least two P1--P5 bands with positive finite flux and error and \({\rm DET\_LIKE}\ge3\), together with finite optical mass and Eddington ratio, but no parent P3 cut.  We repeat exactly the natural-log-flux fit using only P3--P5.  There are 292 finite hard-only fits: 276 use P3/P4, nine use P3/P5 and seven use all three hard bands.  Removing 64 best fits at \(\Gamma_{\rm eff}=0\) leaves 228 objects for the boundary-excluded, same-object comparison.

The boundary exclusion was selected after inspecting the fits, not specified in advance.  For the same 292 boundary-inclusive objects, P1--P5 gives \(\rho=0.400\) and slope 0.578, whereas P3--P5 gives 0.071 and 0.241.  The retained positive result therefore depends on the boundary exclusion.  Hard-only quality cuts additionally requiring a P4 or P5 likelihood of at least 3, \(1<\Gamma_{\rm eff}<3\) and \(\sigma_\Gamma<0.5\) leave only 12 objects, with hard-only \(\rho=0.063\) and slope \(-0.035\).  A quality cut applied to the P1--P5 fit is not a soft-band-independent check.
For the same 228 boundary-excluded objects, the P1--P5 correlation and slope are 0.433 and 0.530, compared with 0.210 and 0.409 for P3--P5.  The hard-only 95 per cent intervals are \([0.079,0.337]\) for \(\rho\) and \([0.196,0.611]\) for the slope.  These are percentile intervals from 5000 paired object bootstraps with optical and fitted-index point estimates fixed; they exclude measurement errors and counterpart uncertainty.

Three sets of 1000 conditional null realizations retain observed optical quantities, redshifts, flux errors and DET\_LIKE values.  The first fixes \(\Gamma_{\rm true}=1.9\); the second draws an independent Gaussian index with mean 1.9 and dispersion 0.22, clipped to 0.8--3.2.  Their expected band fluxes are \(F_{i,\rm exp}=(F_{\rm P3}/I_{\rm P3})I_i\), where \(I_i=\int_{E_{i,1}}^{E_{i,2}}E^{1-\Gamma_{\rm true}}\,\mathrm{d}E\).  Normalization uses observed positive P3 only for the 292 parents capable of entering the fixed-likelihood hard selection, not for all 1695 objects.  Independent Gaussian noise with the observed linear-flux error is added to each band.  Each realization reapplies positive-flux, fixed-likelihood, minimum-band and boundary cuts and repeats the same hard fit.  Errors and likelihoods are not regenerated from photon counts; the simulated retained sample sizes are allowed to change.

The third null retains observed hard indices and permutes optical Eddington ratio within adaptive strata of redshift, log P3 flux and P3 flux/error (minimum stratum size eight).  It preserves local selection structure rather than imposing a universal spectrum.  The positive-tail probability is \((1+n[\rho_{\rm mock}\ge\rho_{\rm obs}])/(1000+1)\).  The positive-tail probabilities are 0.0130 for constant shape, 0.0080 for independent shape scatter, and 0.3856 for stratified permutation.  This contrast shows that the evidence depends on the null assumptions.  These are conditional pipeline tests, not end-to-end eRASS detection simulations.

Most retained fits rely on two hard bands, and the hard-only likelihood information is weak.  Catalogue ECF assumptions, absorption, soft excess, boundary selection and measurement noise can all affect the index.  Neither the smaller hard-band correlation nor the null results isolate soft excess or establish a response-corrected coronal photon-index relation.

\subsection{Fixed bolometric-correction factors}
\label{app:m3_kbol}
We use \(k_X=20\) and \(k_{5100}=9\), fixed across the relevant objects rather than dependent on luminosity or Eddington ratio.  On identical objects,
\begin{equation}
\Delta\log\lambda_{\rm Edd,X}=\log(k_X/20),
\label{eq:fixed_kx_shift}
\end{equation}
so a fixed change in \(k_X\) only translates the horizontal axis.  For the same 2032 primary Class A objects, \(k_X=10,20,30\) gives identical \(\rho=-0.524\) and slope \(-0.204\).  Relative to \(k_X=20\), the factors 10 and 30 shift the horizontal axis by \(-0.301\) and \(+0.176\) dex, respectively.  The original estimator-comparison figure retains its separate 1200-object quality selection.  This invariance is algebraic, not additional evidence for a physical sign reversal.

The optical factor nine is a conventional approximation \citep{Kaspi2000}.  Intrinsic SED diversity, unobserved UV emission and non-simultaneity remain limitations \citep{Richards2006}; neither a 0.30 dex optical-inference term nor a 0.35 dex horizontal-axis stress test measures an uncertainty in this factor.  The DESI DR1 AGN/Galaxy Classification VAC supplies catalogue classification and line information, not object-specific full-SED bolometric corrections for this analysis.  Our team's DESI DR2 full-SED work is ongoing and may provide such corrections in future; it has not been used here.
\endgroup

\begingroup
\section{Matching-radius sensitivity}
\label{app:matching_radius}
The primary analysis retains the 10 arcsec cross-matching radius.  We repeat descriptive trend measurements after reducing only this radius to 7 arcsec.  Applying the radius cut before closest-counterpart deduplication gives exactly the same source identities as restricting the frozen 10 arcsec catalogue.  The unique-pair parent decreases from 3511 to 2684.  We hold spectral measurements, adopted luminosity and mass prescriptions, and all other selection criteria fixed.  The P3 analysis requires positive flux and finite derived quantities, without an additional P3 likelihood cut.  We report Spearman correlations and unweighted linear slopes with a free intercept, using saved point estimates.

For Class A, the optical \(\alpha_{\rm OX}\)--Eddington-ratio sample decreases from 2032 to 1693 objects; \(\rho\) changes from 0.339 to 0.348 and slope from 0.140 to 0.141.  Class B decreases from 410 to 276 objects, with \(\rho=0.636\) versus 0.643 and slopes 0.200 versus 0.213.  The positive optical trends are therefore retained.  Using the same P3-derived X-ray luminosities, fixed \(\Gamma_X=1.8\), and \(k_X=20\), the Class A X-ray-based branch remains negative.  Its correlation changes from \(-0.524\) to \(-0.504\), and its slope from \(-0.204\) to \(-0.195\).

The full-redshift CL-AGN context sample decreases from 78 to 67 usable objects, with optical \(\rho=0.412\) versus 0.403 and slopes 0.199 versus 0.214.  In the homogeneous redshift window, 12 versus ten objects give \(\rho=0.510\) versus 0.503 and slopes 0.241 versus 0.244.  This small comparison subset remains insufficient for an independent population claim.

The independent Class A P1--P5 index sample decreases from 1695 to 1446 objects, retaining a positive optical-Eddington-ratio association.  Its \(\rho\) changes from 0.176 to 0.192 and slope from 0.296 to 0.307.  For the boundary-excluded P3--P5 sample, 228 versus 212 objects give \(\rho=0.210\) versus 0.183 and slopes 0.409 versus 0.374.  These comparisons reuse the original valid-band and boundary selections without refitting spectra.  The principal trend directions survive the radius change, but this descriptive check does not establish statistical equivalence or remove the hard-index limitations discussed in Appendix~\ref{app:m2_hard}.  It measures sensitivity to matching radius, not the false-match fraction or the reliability of individual counterparts.
\endgroup

\bsp	
\label{lastpage}
\end{document}